\documentclass[aps,pra,showpacs,twocolumn]{revtex4-1}
\usepackage{amssymb}
\usepackage{amsmath}
\usepackage{graphicx}
\usepackage{epsfig}
\usepackage{subfigure}
\usepackage{amsfonts}
\usepackage{color}

\begin{document}

\title{Topological phase in Kitaev chain with spatially separated pairing
processes}
\author{Y. B. Shi}
\author{Z. Song}
\email{songtc@nankai.edu.cn}

\begin{abstract}
The dynamic balance between pair creation and annihilation processes takes a
crucial role to the topological superconductivity in Kitaev model. Here we
study the effect of spatial separation of creation and annihilation terms,
i.e., sources and drains of pair are arranged alternatively. In this regard,
a non-Hermitian Hamiltonian is naturally considered, which may possess
complex energy branches. However, when the Bardeen--Cooper--Schrieffer pair
(BCS)-like pair excitation is only considered, the spectrum in such a
subspace is fully real. In particular, the Zak phases extracted from the
pair wave function are quantized and therefore able to characterize the
different phases regardless of the breaking of time reversal symmetry. For
open chain system, the corresponding Majorana lattice is investigated. We
find that although there are complex modes, all the edge modes have zero
energy and obey the bulk-boundary correspondence. This results in the
Kramer-like degeneracy of both the real and complex levels as a signature of
the topologically non-trivial phase.
\end{abstract}

\maketitle
\affiliation{School of Physics, Nankai University, Tianjin 300071, China }

%\affiliation{${}^1$School of Physics, Nankai University, Tianjin 300071, China \\
%${}^2$College of Physics and Materials Science, Tianjin Normal University,
%Tianjin 300387, China}

%%%{11.30.Er Charge conjugation, parity, time reversal, and other discrete symmetries}
%%%{75.10.Pq Spin chain models}
%%%{75.10.Jm Quantized spin models, including quantum spin frustration}
%%%{64.70.Tg Quantum phase transitions}
%%%{64.60.fd General theory of critical region behavior}

\section{Introduction}

{In traditional quantum mechanics, the fundamental postulate of
the Hermiticity of the Hamiltonian ensures\ the reality of the spectrum and
the unitary dynamics for a closed quantum system \cite{DAMQ}. In general,
any Hermitian Hamiltonian can also be decomposed into two non-Hermitian
sub-Hamiltonians which are Hermitian conjugation of each other. There are
many different ways of decomposing a Hamiltonian. In this sense, the reality
of spectrum roots in the balance of the actions of the two non-Hermitian
sub-Hamiltonians. Intuitively, the Hermiticity is not necessary for the
balance. In fact, this postulate has been questioned by the existence of
entirely real spectra of a certain class of non-Hermitian Hamiltonians \cite%
{Bender,Bender1}. Such kinds of Hamiltonians, referred to as
pseudo-Hermitian operator \cite{Ali1,Ali2,Ali3,Ali4,Ali5}, possess a
particular symmetry, i.e., it commutes with the combined operator $\mathcal{%
PT}$, but not necessarily with $\mathcal{P}$\ and $\mathcal{T}$\
separately.\ Here $\mathcal{P}$\ is a unitary operator, such as parity,
translation, rotation operators etc., while $\mathcal{T}$\ is an
anti-unitary operator, such as time-reversal operator. The combined symmetry
is said to be unbroken if every eigenstate of the Hamiltonian is $\mathcal{PT%
}$-symmetric and the entire spectrum is real; it is said to be spontaneously
broken if some eigenstates of the Hamiltonian are not the eigenstates of the
combined symmetry. Such a symmetry can be regarded as a demonstration of the
balance, for instance, a simple gain--loss-balanced system in Refs. \cite%
{Jin1,Jin2,Jin3}. This concept should be extended to a system without
conservation of particle number. }

\begin{figure}[t]
\centering \includegraphics[width=0.5\textwidth]{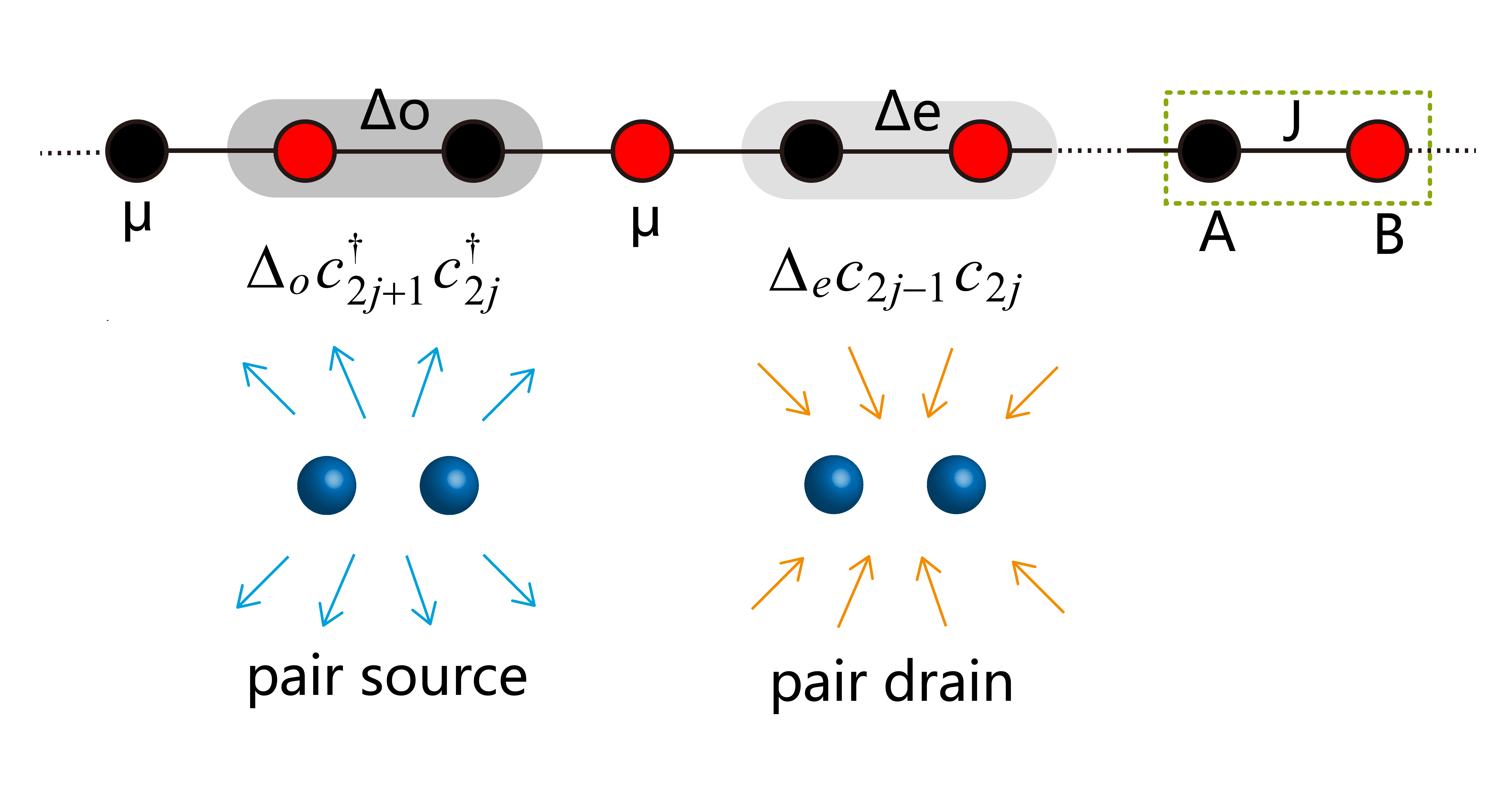} 
\caption{(a) Schematic of the 1D Kitaev model for spinless fermion with
non-Hermitian imbalanced pair terms, arising from specially separated
creation and annihilation. Here $J$ is the hopping strength between two
adjacent sites and the dark (light) shade region represents the pair
creation (annihilation) on odd (even) dimer with amplitude $\Delta _{\mathrm{o}}$ $(\Delta _{\mathrm{e}})$. The staggered pair terms describe the source
and drain of nearest neighboring fermion pairs. The unit cell consists of
two sublattices A and B in black and red as indicated inside the dashed
green rectangle. $\protect\mu $ is the chemical potential.} \label{fig1}
\end{figure}

{The Kitaev model is a typical one for violating particle
conservation. It is believed to capture the feature of one-dimensional
topological superconductor for spinless fermions with $p$-wave pairing \cite%
{Kitaev}. The topological superconducting phase has been demonstrated by
unpaired Majorana modes exponentially localized at the ends of open Kitaev
chains \cite{Sarma,Stern,Alicea}. Intimately related to the superconducting
phase,\ the dynamic balance between pair creation and annihilation processes
takes a crucial role, which violates the conservation of the fermion number
but preserves its parity. It is the fermionized version of the familiar
one-dimensional transverse-field Ising model \cite{Pfeuty}, which is one of
the simplest solvable models exhibiting quantum criticality and
demonstrating a quantum phase transition with spontaneous symmetry breaking 
\cite{SachdevBook}. So far, most of the investigations on the Kitaev model
have focused on the case with the balanced pair creation and annihilation at
the same location, as pair source and drain, respectively. In principle,
there are many other types of balance between pair creation and
annihilation\ such as the sources and drains of pair are arranged
alternatively (Fig. \ref{fig1}). Intuitively, such kinds of balance should
take the similar role in the existence of the gapped superconducting phase.
However, when these cases are considered, the traditional method is no
longer valid because a non-Hermitian Hamiltonian is involved.}

The strategy of this paper is to study the topological phase in Kitaev chain
with spatially separated pairing processes based on the non-Hermitian
quantum theory. Non-Hermitian Kitaev models \cite%
{Law,Tong,Yuce,You,Klett,Menke,LCPRB,YXMPRA,DMTN} and Ising models \cite%
{ZXZ,LCPRA90,LCPRA92,LCPRA94} has been studied within the pseudo-Hermitian
framework. Also, the experimental schemes for realizing the Kitaev model and
related non-Hermitian systems have been presented in Refs. \cite{Franz} and 
\cite{Ueda}, respectively. In this work, we study the effect of spatial
separation of creation and annihilation terms, i.e., sources and drains of
pair are arranged alternatively. It is a non-Hermitian model with
time-reversal symmetry, rather than a combined $\mathcal{PT}$\ symmetry.
Exact solution shows a rich phase diagram, including gapped superconducting
phases associated with both time-reversal symmetry and broken time-reversal
symmetry. However, when\ only the Bardeen--Cooper--Schrieffer pair
(BCS)-like excitation is considered, the spectrum in the subspace is fully
real. There are two topological superconducting phases characterized by the
extended Zak phase, which is defined in terms of biorthonormal inner product 
\cite{Raghu,Shindou}. The Zak phases extracted from the pair wave function
are quantized and therefore able to characterize the different phases
regardless of the breaking of time reversal symmetry. We also investigate
the bulk-edge correspondence in such a non-Hermitian system\ via\ the
Majorana transformation. The corresponding Majorana lattice for open chain
system is investigated. We find that\ although there are complex modes, all
the edge modes have zero energy and obey the bulk-edge correspondence. This
results in the Kramer-like degeneracy of both the real and complex levels as
a signature of the topologically non-trivial phase.

{In comparison with the original Kitaev chain, the present
non-Hermitian Kitaev chain remains the topological phase, which can be
identified by the Zak phase, two-types of order parameters, in the context
of biorthonormal inner product. In addition, the bulk-boundary
correspondence still holds although the Majorana lattice is non-Hermitian.
However, on the other hand, the spectrum becomes complex in the symmetry
breaking region, but keeping the reality of the pair spectrum. We also find
that the phase boundary is independent of the symmetry breaking region, but
dependent of the strength of the pair term.}

This paper is organized as follows. In Section \ref{Model and pair spectrum}%
, we describe the model Hamiltonian and present the solutions. In Section %
\ref{Phase diagram and order parameter}, we investigate the phase diagram
and analyze the signatures of the phase boundary based on the solutions. In
Section \ref{Topological phase} we construct the pair wave function and
calculate the Zak phase to discuss the topological properties of the system.
In Section \ref{Bulk-boundary correspondence}, we study the corresponding
Majorana lattice to show the existence of the bulk-edge correspondence. In
Section \ref{Kramer-like degeneracy} we provide the implication of the zero
modes. Finally, we give a summary and discussion in Section \ref{sec_summary}%
. Some details of derivations are placed in the Appendix.

\section{Model and pair spectrum}

\label{Model and pair spectrum}

We consider the following fermionic Hamiltonian on a lattice of length $2N$

\begin{eqnarray}
H &=&\sum\limits_{l=1}^{2N}[Jc_{l}^{\dag }c_{l+1}+\text{\textrm{H.c.}}+\mu
\left( 1-2n_{l}\right) ]  \notag \\
&&+\sum\limits_{j=1}^{N}(e^{i\varphi }\Delta _{\mathrm{o}}c_{2j+1}^{\dagger
}c_{2j}^{\dagger }+e^{-i\varphi }\Delta _{\mathrm{e}}c_{2j-1}c_{2j}),
\label{H}
\end{eqnarray}%
where $c_{l}^{\dag }$ $(c_{l})$\ is a fermionic creation (annihilation)
operator on site $l$, $n_{l}=c_{l}^{\dag }c_{l}$, $J$ the tunneling rate, $%
%TCIMACRO{\U{3bc} }%
%BeginExpansion
\mu
%EndExpansion
$ the chemical potential, and real number $\Delta _{\mathrm{o}}$ $(\Delta _{%
\mathrm{e}})$\ the strength of the $p$-wave pair creation (annihilation) on
odd (even) dimer. For a closed chain, we define $c_{2N+1}=c_{1}$ and for an
open chain, we set $c_{2N+1}=0$. The Kitaev model is known to have a rich
phase diagram in its Hermitian version, i.e.,\textbf{\ }$H\rightarrow
H+H^{\dag }$\textbf{. }

Comparing with the non-Hermitian Kitaev model in previous works \cite%
{Law,Tong,Yuce,You,Klett,Menke,LCPRB,YXMPRA,DMTN},\textbf{\ }the present
model has time-reversal symmetry and its non-Hermiticity arises from the
staggered unidirectional pairing term, which has never been reported in the
literatures. The physical intuition for this modification is simple. It is
the simplest model to characterize the system with non-locally pairing
processes, i.e., the pair creation and annihilation occurring at different
dimers.

In this work, we only consider the reduced Hamiltonian by taking $\Delta _{%
\mathrm{o}}=\Delta _{\mathrm{e}}=\Delta >0$ and $\varphi =0$.
Mathematically, the reduced and the original Hamiltonians can be mapped from
one to another via a local transformation. Actually, taking the similarity
transformation

\begin{equation}
c_{j}=e^{-\gamma /2}e^{-i\varphi /2}c_{j},\overline{c}_{j}=e^{\gamma
/2}e^{i\varphi /2}c_{j}^{\dag },
\end{equation}%
with $e^{\gamma }=\Delta _{\mathrm{o}}/\Delta _{\mathrm{e}}$ and $\Delta =%
\sqrt{\Delta _{\mathrm{o}}\Delta _{\mathrm{e}}}$, one can get the reduced
Hamiltonian with the operators $H\left( c_{j},\overline{c}_{j}\right) $.
Note that operators $\left( c_{j},\overline{c}_{j}\right) $\ satisfy the
canonical relations%
\begin{equation}
\left\{ c_{i},\overline{c}_{j}\right\} =\delta _{ij},\left\{
c_{i},c_{j}\right\} =\left\{ \overline{c}_{i},\overline{c}_{j}\right\} =0%
\text{,}
\end{equation}%
and than any two Hamiltonians $H\left( c_{i},\overline{c}_{j}\right) $ and $%
H\left( c_{i},c_{j}^{\dag }\right) $\ have the same structure of the
solution \cite{LCPRB}, sharing similar features.

Before solving the Hamiltonian (reduced Hamiltonian with $\Delta _{\mathrm{o}%
}=\Delta _{\mathrm{e}}=\Delta >0$ and $\varphi =0$), it is profitable to
investigate the symmetry of the system and its spontaneously breaking in the
eigenstates. It can be checked that the model respects two evident
symmetries. The first one is the parity symmetry%
\begin{equation}
\left[ \Pi ,H\right] =0,
\end{equation}%
where the parity operator about fermion number is defined as $\Pi
=\prod_{j=1}^{N}(-1)^{n_{j}}$. Secondly, by the direct derivation, we have $%
\left[ \mathcal{T},H\right] =0$, i.e., the Hamiltonian is a time reversal ($%
\mathcal{T}$) invariant, where the antilinear time reversal operator $%
\mathcal{T}$ has the function $\mathcal{T}i\mathcal{T=-}i$. In general, the $%
\mathcal{T}$\ symmetry in a non-Hermitian model plays the same role as the $%
\mathcal{PT}$\ symmetry, probably resulting\ a $\mathcal{PT}$\
pseudo-Hermitian system. It motivates a more systematic study of such a
model, including the influence of the spontaneously breaking to the
topological phase\textbf{.}

\begin{figure}[t]
\centering\includegraphics[width=0.47\textwidth]{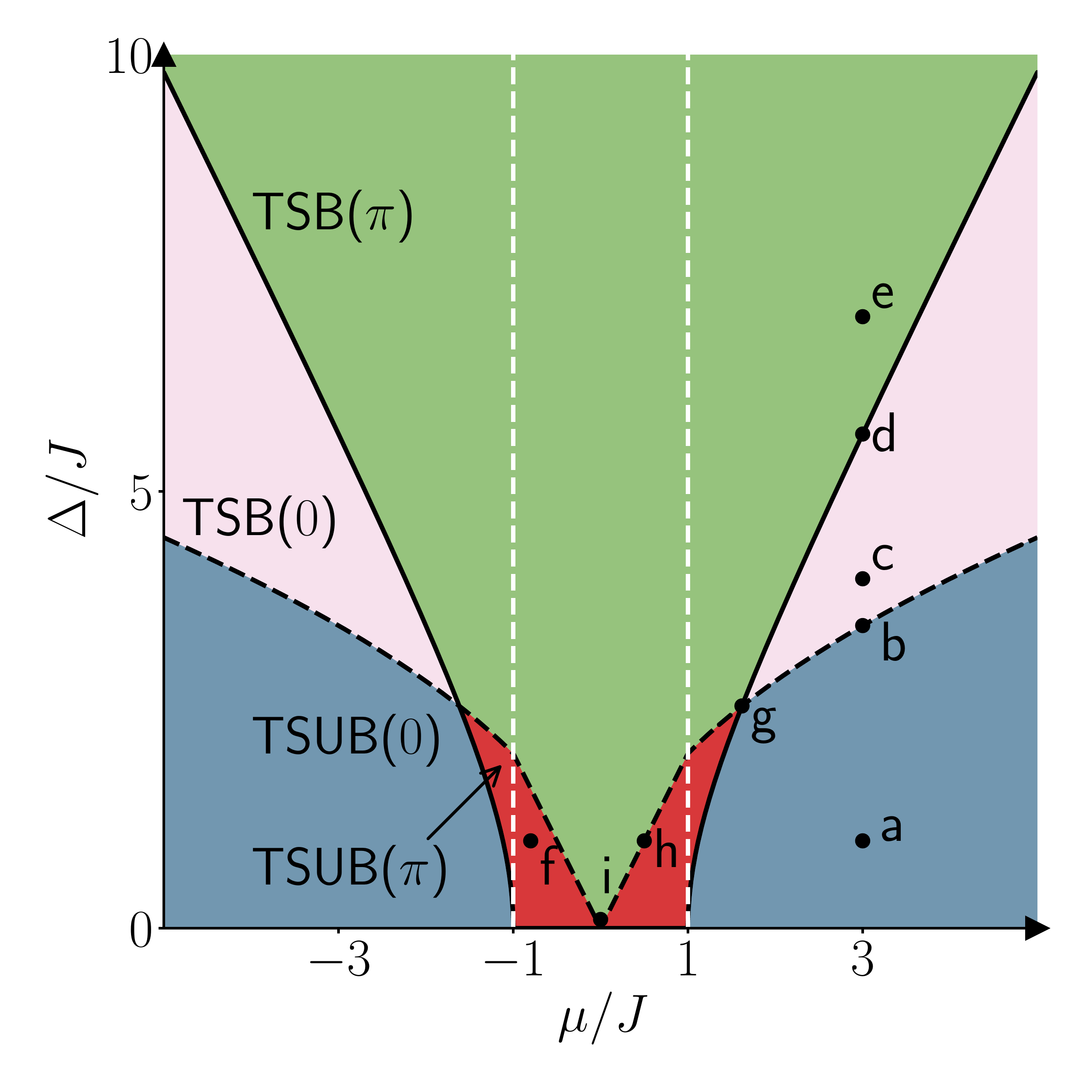}  
\caption{{Phase diagram for a closed chain in $\protect\mu -\Delta $ parameter plane. The black solid curves indicate the phase
			boundary, which also separates topological phases identified by the pair Zak
			phase defined in Eq. (\protect\ref{Z_sum}). Phase boundary is identified by
			by zero energy gap in Eq. (\protect\ref{phase boundary}). In the non-trivial
			regions, The Kramer-like degeneracy exists. The reality of the spectrum is
			differentiated by the black dashed curves, whose form is provided in Eq. (\protect\ref{broken}). It is also the boundary between the $\mathcal{T}$-symmetry-broken and -unbroken regions. (a-h) denote several typical points
			in different regions and boundaries, which will be investigated in Fig. 
			\protect\ref{fig3}. The phase boundary of the traditional Kitaev chain is
			indicated by the white dashed lines.} } \label{fig2}
\end{figure}

We introduce the Fourier transformations in two sub-lattices{
\begin{equation}
c_{j}=\frac{1}{\sqrt{N}}\sum_{k}e^{ikn}\left\{ 
\begin{array}{cc}
\alpha _{k}, & j=2n \\ 
\beta _{k}, & j=2n-1%
\end{array}%
\right. ,
\end{equation}%
}where $n=1,2,...,N$, $k=2m\pi /N$, $m=0,1,2,...,N-1$. Inversely, the
spinless fermionic operators in $k$ space $\alpha _{k},$\ $\beta _{k}$ are%
\begin{equation}
\left\{ 
\begin{array}{cc}
\alpha _{k}=\frac{1}{\sqrt{N}}\sum\limits_{j}e^{-ikn}c_{j}, & j=2n \\ 
\beta _{k}=\frac{1}{\sqrt{N}}\sum\limits_{j}e^{-ikn}c_{j}, & j=2n-1%
\end{array}%
\right. .
\end{equation}%
The Hamiltonian with periodic boundary condition can be block diagonalized
by this transformation due to its translational symmetry, i.e.,%
\begin{equation}
H=\sum_{k\in (0,\pi ]}H_{k}=\psi _{k}^{\dagger }h_{k}\psi _{k},
\end{equation}%
satisfying $\left[ H_{k},H_{k^{\prime }}\right] =0$, where the operator
vector $\psi _{k}^{\dagger }=\left( 
\begin{array}{cccc}
\alpha _{k}^{\dagger } & \beta _{k}^{\dagger } & \alpha _{-k} & \beta _{-k}%
\end{array}%
\right) $, and the core matrix%
\begin{equation}
h_{k}=\left( 
\begin{array}{cccc}
-2\mu & \gamma _{k} & 0 & -\Delta e^{ik} \\ 
\gamma _{k}^{\ast } & -2\mu & \Delta e^{-ik} & 0 \\ 
0 & -\Delta & 2\mu & -\gamma _{k} \\ 
\Delta & 0 & -\gamma _{k}^{\ast } & 2\mu%
\end{array}%
\right) ,  \label{h}
\end{equation}%
with $\gamma _{k}=J\left( 1+e^{ik}\right) $. So far the procedure is the
same as that for solving a Hermitian Hamiltonian. Obviously, matrix $h_{k}$\
is no longer Hermitian. To diagonalize a non-Hermitian Hamiltonian, one
should introduce the Bogoliubov transformation in the complex version:%
\begin{equation}
A_{\rho \sigma }^{k}=\left( V^{k}\psi _{k}\right) _{\rho \sigma },\overline{A%
}_{\rho \sigma }^{k}=(\psi _{k}^{\dagger }\overline{V}^{k})_{\rho \sigma },
\end{equation}%
with $\rho ,\sigma =\pm $, where $V^{k}$ and $\overline{V}^{k}=\left(
V^{k}\right) ^{-1}$ are similarity transformation for the diagonalization of 
$h_{k}$, i.e.,%
\begin{equation}
V^{k}h_{k}\overline{V}^{k}=\mathrm{dia}\left( 
\begin{array}{cccc}
\varepsilon _{++}^{k} & \varepsilon _{+-}^{k} & \varepsilon _{--}^{k} & 
\varepsilon _{-+}^{k}%
\end{array}%
\right) .
\end{equation}%
In the Appendix \ref{AppendixA}\textbf{,} the explicit form of $V^{k}$ and $%
\overline{V}^{k}$ are presented. This procedure essentially establishes the
biorthogonal basis, which is a crucial step to solve a pseudo-Hermitian
Hamiltonian. Obviously, complex Bogoliubov modes $(A_{\rho \sigma }^{k},%
\overline{A}_{\rho \sigma }^{k})$ satisfy the anticommutation relations%
\begin{eqnarray}
\{A_{\rho \sigma }^{k},\overline{A}_{\rho ^{\prime }\sigma ^{\prime
}}^{k^{\prime }}\} &=&\delta _{kk^{\prime }}\delta _{\rho \rho ^{\prime
}}\delta _{\sigma \sigma ^{\prime }}, \\
\{A_{\rho \sigma }^{k},A_{\rho ^{\prime }\sigma ^{\prime }}^{k^{\prime }}\}
&=&\{\overline{A}_{\rho \sigma }^{k},\overline{A}_{\rho ^{\prime }\sigma
^{\prime }}^{k^{\prime }}\}=0,
\end{eqnarray}%
which result in the diagonal form of the Hamiltonian%
\begin{equation}
H=\sum\limits_{k}\sum\limits_{\rho ,\sigma =\pm }\varepsilon _{\rho \sigma
}^{k}\overline{A}_{\rho \sigma }^{k}A_{\rho \sigma }^{k}.
\end{equation}%
Here the dispersion relation of the quasiparticle is given by%
\begin{equation}
\varepsilon _{\rho \sigma }^{k}=\rho \sqrt{\Lambda _{k}+\sigma \sqrt{\Omega
_{k}}},  \label{1E}
\end{equation}%
where%
\begin{eqnarray}
\Lambda _{k} &=&4\mu ^{2}+4J^{2}\cos ^{2}\frac{k}{2}-\Delta ^{2}\cos k, \\
\Omega _{k} &=&64\mu ^{2}J^{2}\cos ^{2}\frac{k}{2}-16J^{2}\Delta ^{2}\cos
^{4}\frac{k}{2}  \notag \\
&&-\Delta ^{4}\sin ^{2}k.
\end{eqnarray}%
It can been shown that\textbf{\ }$\mathcal{T}\overline{A}_{\rho \sigma
}^{k}A_{\rho \sigma }^{k}\mathcal{T}^{-1}=\overline{A}_{\rho \sigma
}^{k}A_{\rho \sigma }^{k}$ if $\varepsilon _{\rho \sigma }^{k}$\ is real,
while $\mathcal{T}\overline{A}_{\rho \sigma }^{k}A_{\rho \sigma }^{k}%
\mathcal{T}^{-1}=\overline{A}_{\rho \overline{\sigma }}^{k}A_{\rho \overline{%
\sigma }}^{k}\neq \overline{A}_{\rho \sigma }^{k}A_{\rho \sigma }^{k}$\
(with $\overline{\sigma }=-\sigma $ and $\overline{\rho }=-\rho $) when $%
\varepsilon _{\rho \sigma }^{k}$\ becomes complex, which indicate the
symmetry unbroken and broken regions. The boundary of two regions in the $%
\Delta -\mu $\ plane (in unit of $J$) is%
\begin{equation}
\Delta =\left\{ 
\begin{array}{cc}
2\mu , & \left\vert \mu \right\vert <J \\ 
2\sqrt{\left\vert \mu \right\vert J} & \left\vert \mu \right\vert >J%
\end{array}%
\right. ,  \label{broken}
\end{equation}%
which is plotted in Fig. \ref{fig2} by the black dashed curves.

However, when we consider a pair of Bogoliubov modes,\textbf{\ }we find that%
\textbf{\ }$\mathcal{T}\overline{A}_{\rho \overline{\sigma }}^{k}\overline{A}%
_{\rho \sigma }^{k}A_{\rho \overline{\sigma }}^{k}A_{\rho \sigma }^{k}%
\mathcal{T}^{-1}$ $=$ $\overline{A}_{\rho \overline{\sigma }}^{k}\overline{A}%
_{\rho \sigma }^{k}A_{\rho \overline{\sigma }}^{k}A_{\rho \sigma }^{k}$%
\textbf{\ }is always true. Accordingly, the pair energy%
\begin{equation}
E_{\rho }^{k}=\varepsilon _{\rho +}^{k}+\varepsilon _{\rho -}^{k}=\rho \sqrt{%
\Lambda _{k}+r_{k}},  \label{2E}
\end{equation}%
referred to as pair spectrum, is always real with%
\begin{eqnarray}
r_{k} &=&\sqrt{x_{k}^{2}+y_{k}^{2}}, \\
x_{k} &=&(2J^{2}+\Delta ^{2})\cos k+2J^{2}-4\mu ^{2}, \\
y_{k} &=&\Delta \sqrt{\Delta ^{2}+4J^{2}}\sin k.
\end{eqnarray}%
The expression $r_{k}\left( x_{k},y_{k}\right) $\ is crucial for the
discussion about topological phase in the following sections.

Then any $N_{l}$-quasiparticle eigenstate of the Hamiltonian $H$\ can be
written in the form%
\begin{equation}
\prod\limits_{l=1}^{N_{l}}\overline{A}_{\rho _{l}\sigma
_{l}}^{k_{l}}\left\vert \text{Vac}\right\rangle ,
\end{equation}%
with eigen energy $\sum\nolimits_{l=1}^{N_{l}}\varepsilon _{\rho _{l}\sigma
_{l}}^{k_{l}}$, which associates with the biorthogonal eigenstate%
\begin{equation}
\left\langle \overline{\text{Vac}}\right\vert
\prod\limits_{l=1}^{N_{l}}A_{\rho _{l}\sigma _{l}}^{k_{l}}
\end{equation}%
of the Hamiltonian $H^{\dag }$ with eigen energy $\sum%
\nolimits_{l=1}^{N_{l}}\left( \varepsilon _{\rho _{l}\sigma
_{l}}^{k_{l}}\right) ^{\ast }$, where $\left\vert \text{Vac}\right\rangle \
(\left\langle \overline{\text{Vac}}\right\vert )$ is the vacuum state of
operator $A_{\rho _{l}\sigma _{l}}^{k_{l}}$ ($\overline{A}_{\rho _{l}\sigma
_{l}}^{k_{l}}$), defined as $A_{\rho _{l}\sigma _{l}}^{k_{l}}\left\vert 
\text{Vac}\right\rangle =0$ ($\left\langle \overline{\text{Vac}}\right\vert 
\overline{A}_{\rho _{l}\sigma _{l}}^{k_{l}}=0$). The many-particle
eigenstate can have complex eigen enrgy in the broken region. Nevertheless,
there is a set of eigenstates which are constructed by a set of pair
operators $\{\overline{A}_{\rho +}^{k}\overline{A}_{\rho -}^{k}\}$, always
possessing real eigen energy in both broken and unbroken regions. We refer
the set of states as BCS-pair states since they contain the pair states $%
\alpha _{k}^{\dagger }\alpha _{-k}^{\dagger }\left\vert 0\right\rangle $, $%
\beta _{k}^{\dagger }\beta _{-k}^{\dagger }\left\vert 0\right\rangle $, $%
\alpha _{k}^{\dagger }\beta _{-k}^{\dagger }\left\vert 0\right\rangle $, and 
$\alpha _{-k}^{\dagger }\beta _{k}^{\dagger }\left\vert 0\right\rangle $,
where $\left\vert 0\right\rangle $\ is the vacuum state of fermion operator $%
c_{j}$. Among them, the one with lowest eigen energy%
\begin{equation}
E_{\mathrm{g}}=\sum_{k>0}E_{-}^{k}
\end{equation}%
is defined as pair ground state%
\begin{equation}
\left\vert \text{G}\right\rangle =\prod_{k>0}\overline{A}_{\rho +}^{k}%
\overline{A}_{\rho -}^{k}\left\vert \text{Vac}\right\rangle .
\end{equation}%
In parallel, we have $\left\langle \overline{\text{G}}\right\vert
=\left\langle \overline{\text{Vac}}\right\vert \prod_{k>0}A_{\rho
-}^{k}A_{\rho +}^{k}$\ with $\langle \overline{\text{G}}\left\vert \text{G}%
\right\rangle =1$, satisfying $\left\langle \overline{\text{G}}\right\vert
H=\left\langle \overline{\text{G}}\right\vert E_{\mathrm{g}}$.

\begin{figure*}[t]
\centering \includegraphics[width=0.8\textwidth]{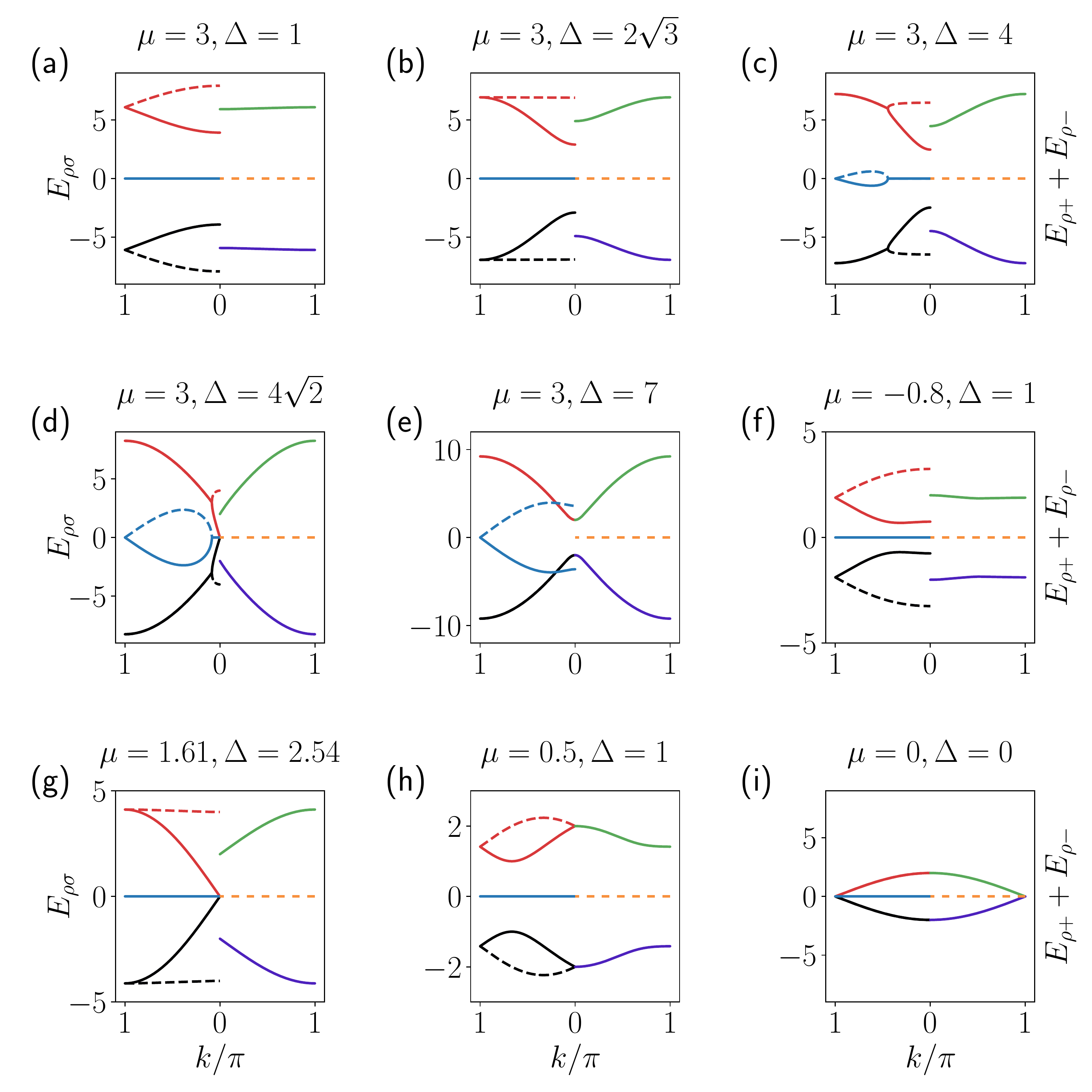}  
\caption{{Energy spectra from Eqs. (\protect\ref{1E}) and ( 
\protect\ref{2E}) for the Hamiltonians at typical points in different
regions of Fig. \protect\ref{fig2}. The red dashed (solid) line stands for $%
\mathrm{Re}(\protect\varepsilon _{++}^{k})$\ ($\mathrm{Re}(\protect%
\varepsilon _{+-}^{k})$), and the black dashed (solid) line stands for Re($%
\protect\varepsilon _{-+}^{k}$)\ (Re($\protect\varepsilon _{--}^{k}$)). The
blue dashed (solid) line stands for Im($\protect\varepsilon _{--}^{k}$)\
(Im( $\protect\varepsilon _{-+}^{k}$)). Similarly, the green (purple) line
stands for Re($\protect\varepsilon _{++}^{k}+\protect\varepsilon _{+-}^{k}$)$%
/2$\ (Re($\protect\varepsilon _{-+}^{k}+\protect\varepsilon _{--}^{k}$)$/2$%
), and the orange line stands for Im($\protect\varepsilon _{--}^{k}+\protect%
\varepsilon _{-+}^{k}$)$/2$. Other parameter $N=1000$.}}
\label{fig3}
\end{figure*}

{In Fig \ref{fig3}, we plot the Energy spectra for the
Hamiltonian at typical points in Fig. \ref{fig2}. The dispersion relation of
the quasiparticle obtained in Eq. (\ref{1E}) is plotted on the left side and
the pair energy is plotted at the right side. The energy gap closes at the
condition $\varepsilon _{- +}^{k}=\varepsilon _{+ -}^{k}$. The solution is 
\begin{equation}
\Delta^{2}+4J^{2}-4\mu^{2}=0.  \label{ZeroGap}
\end{equation}%
In the following several Sections, we will prove that the system experiences
topological phase transition when the closing occurs, which is the main
focus of our investigation. The boundary is plotted in Fig. \ref{fig2} by
the black solid curves. It show that the phase diagram is extended in $\mu $%
\ direction as $\Delta $\ increases comparing to that of traditional Kitaev
chain indicated by the white dashed lines. } {Now, with the help
of Eqs. (\ref{broken}) and (\ref{ZeroGap}), the phase diagram can be
distinguished into four parts. we classified These four regions by the pair
Zak phase $\mathcal{Z}$\ of the ground state, which will be introduced in
detail in Section \ref{Topological phase}, and the time reversal symmetry of
the eigenvectors. They are listed as: (i) Time symmetry is broken with $%
\mathcal{Z}=\pi $\ [TSB($\pi $)]; (ii) Time symmetry is unbroken with $%
\mathcal{Z}=\pi $\ [TSUB($\pi $)]; (iii) Time symmetry is broken with $%
\mathcal{Z}=0$\ [TSB($0$)]; (iv) Time symmetry is unbroken with $\mathcal{Z}%
=0$\ [TSUB($0$)].}

{The spectrum properties of the four phases are summarized in
Table \ref{Table I}, which clearly indicates the distinction between two
phases. In Fig. \ref{fig2}, we mark the different phases with different
colors. Fig. \ref{fig3} reveals these properties straightforwardly. (i) At
the phase boundary, (d) and (g), positive and negative branches touch at $k=0
$. Otherwise there are energy gap. (ii) In TSUB regions, (a) and (f), four
branches are all real. In TSB regions, (c), (d), and (e), all of them become
complex. (iii) At the $\mathcal{T}$ -symmetry-broken boundary, (b), (g), and
(h), the branches are still real. A slight increase of $\Delta $ or decrease
of $\mu $ \ can result in the appearance of complex energy levels from $%
k=\pi $. For the point (i), the model reduces to a non-interacting Hermitian
model. Notably, we note that all curves of $\left( \varepsilon _{\rho
+}^{k}+\varepsilon _{\rho -}^{k}\right) /2$ are real. In addition, the
classification of phases is independent of the boundary conditions. In fact,
it can also be obtained from a open chain since (i) the spectrum of a system
is independent of the boundary conditions large $N$, and (ii) the Zak phase
for a ring system is equivalent to the number of zero modes of Majorana
lattice with open boundary condition, according to the bulk-boundary
correspondence.} In the following, we will investigate the phase diagram
based on the properties of the pair ground state $(\left\vert \text{G}%
\right\rangle ,\left\langle \overline{\text{G}}\right\vert )$.

\begin{table}[tbp]
\caption{{Classification of four quantum phases by the
			pair Zak phase $\mathcal{Z}$ of the ground state, symmetry of eigenvector,
			reality of two kinds of spectrum, energy gap and the edge mode of Majorana
			lattice.}}\textbf{\ \renewcommand\arraystretch{1}}

\begin{tabular}{ccccc}
\hline\hline
& TSB($\pi $) & TSUB($\pi $) & TSB($0$) & TSUB($0$) \\ \hline
Eigenvector symmetry & no & yes & no & yes \\ 
Eigen spectrum & complex & real & complex & real \\ 
Pair spectrum & real & real & real & real \\ 
Pair-spectrum gap & yes & yes & yes & yes \\ 
Edge mode & yes & yes & no & no \\ 
Typical point & e & f & c & a \\ \hline\hline
\end{tabular}
\label{Table I}
\end{table}

\begin{figure*}[t]
\centering \includegraphics[width=0.9\textwidth]{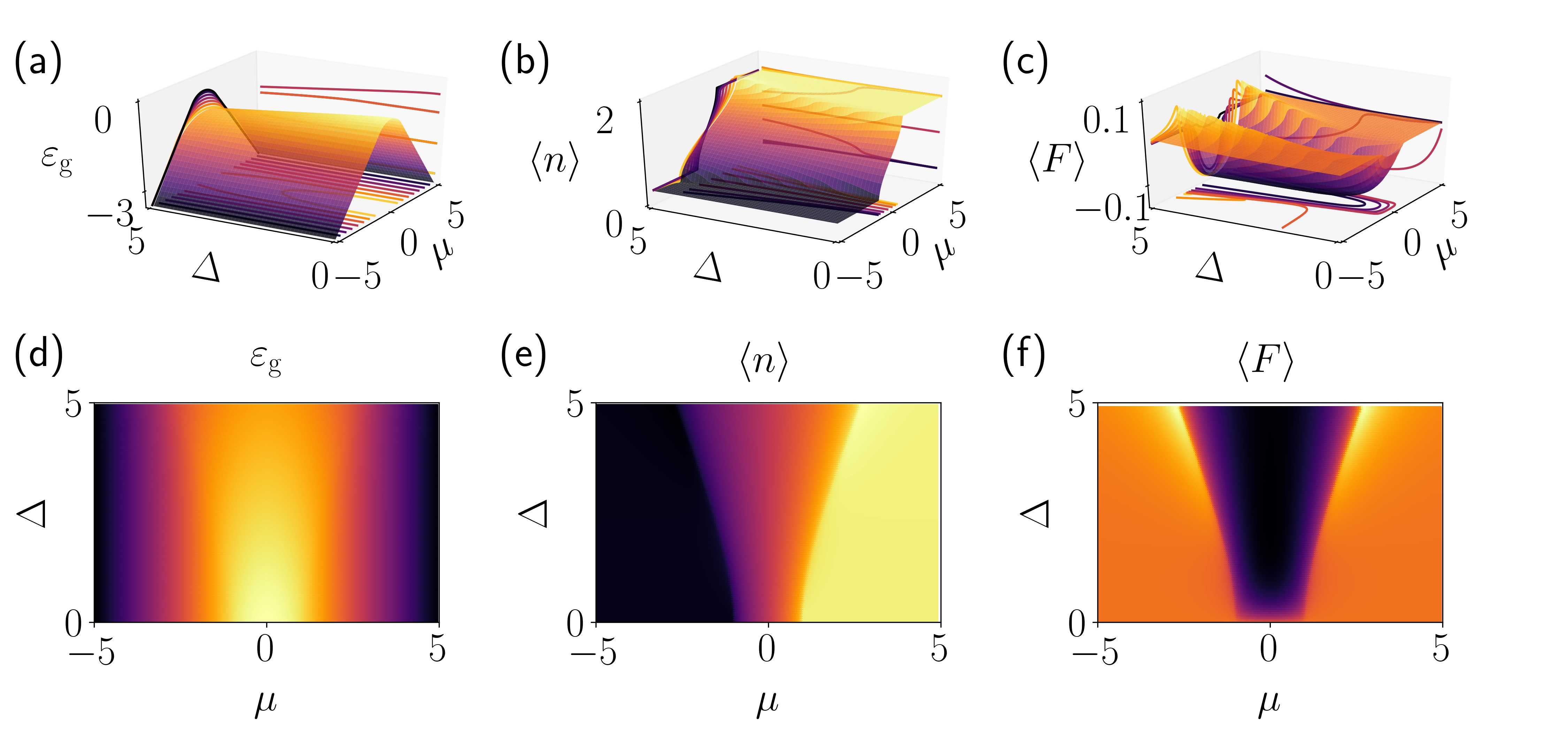}
\caption{3D plots of (a) $\protect\varepsilon _{\mathrm{g}}$, (b) $%
\left\langle n\right\rangle $ and (c) $\left\langle F\right\rangle $
obtained from Eqs. (\protect\ref{GED}), (\protect\ref{OP}) and (\protect\ref%
{APD}), respectively. (d)-(f) are the corresponding top views of them. It
can be seen that $\left\langle n\right\rangle $ and $\left\langle
F\right\rangle $ are discontinuous at the phase boundary. Other parameter is 
$N=4000$.}
\label{fig4}
\end{figure*}

\section{Phase diagram and order parameter}

\label{Phase diagram and order parameter}

In general, a quantum phase transition for a Hermitian system occurs when
the ground state experiences sudden change \cite{SachdevBook}. The phase
boundary in parameter space is characterized by the opening or closing of
the energy gap between ground and first excited states. Remarkably, it
always associated with discontinuity of the derivative of density of
groundstate energy, or other observables, and even the change of topological
index for topological quantum phase transition. It is worthy to investigate
the present system systematically due to its non-Hermiticity. Two main
questions should be addressed: (i) Is there a conventional quantum phase
transition for the pair ground state of such a non-Hermitian system? (ii)
What happens in the pair ground state at symmetric broken boundary?

According to the theory of quantum phase transition in a Hermitian system,
the quantum phase boundary locates at the degenerate point, or gap closing
point. For the present system, it corresponds to $\varepsilon _{\rho
-}^{k_{c}}=0$, which determines the phase boundary in the parameter space.
From equation $\varepsilon _{\rho -}^{k_{c}}=0$,\ we have $r_{k_{c}}=\sqrt{%
x_{k_{c}}^{2}+y_{k_{c}}^{2}}=0$, or explicitly%
\begin{eqnarray}
(2J_{c}^{2}+\Delta _{c}^{2})\cos k_{c}+2J_{c}^{2}-4\mu _{c}^{2} &=&0,
\label{EG1} \\
\sqrt{\Delta _{c}^{4}+4\Delta _{c}^{2}J_{c}^{2}}\sin k_{c} &=&0.  \label{EG2}
\end{eqnarray}%
The solution is $k_{c}=0$, resulting in the phase boundary%
\begin{equation}
\Delta _{c}^{2}+4J_{c}^{2}-4\mu _{c}^{2}=0,  \label{phase boundary}
\end{equation}%
which is plotted in Fig. \ref{fig2} by the black solid curves. We note that
the quantum phase transition occurs at the zero point of $r_{k}=\sqrt{%
x_{k}^{2}+y_{k}^{2}}$, which is crucial for the investigation of topological
quantum phase transition in the next section. When taking $\Delta _{c}=0$,
the phase boundary reduces to $\mu _{c}=\pm J_{c}$ (white dashed lines in
Fig. \ref{fig2}), which is the one for the standard Hermitian Kitaev chain.
In this regard, the present phase diagram is extended in $\mu $ direction as 
$\Delta $ increases comparing to that of traditional Kitaev chain.

Now we consider the behavior of the density of groundstate energy\textbf{\ }$%
\varepsilon _{\mathrm{g}}$\textbf{. }In thermodynamic limit, $N\rightarrow
\infty $, we have

\begin{equation}
\varepsilon _{\mathrm{g}}=\frac{E_{\mathrm{g}}}{2N}=\frac{1}{2\pi }%
\int_{0}^{\pi }E_{-}^{k}\mathrm{d}k.  \label{GED}
\end{equation}%
We note that the integrand $E_{-}^{k}$\ at $k=0^{+}$%
\begin{equation}
E_{-}^{0^{+}}=-\sqrt{4\mu ^{2}+4J^{2}-\Delta ^{2}+\left\vert
x_{0^{+}}\right\vert }
\end{equation}%
contains a term%
\begin{equation}
\left\vert x_{0^{+}}\right\vert =\left\vert \Delta ^{2}+4J^{2}-4\mu
^{2}\right\vert ,
\end{equation}%
which is non-analytic at the boundary, i.e., $\left\vert
x_{0^{+}}\right\vert =0$, resulting non-analytic $\varepsilon _{\mathrm{g}}$%
. Accordingly, it indicates the discontinuity of $\left( \partial
\varepsilon _{\mathrm{g}}/\partial \Delta ,\partial \varepsilon _{\mathrm{g}%
}/\partial \mu \right) $\ and divergence of $\left( \partial ^{2}\varepsilon
_{\mathrm{g}}/\partial \Delta ^{2},\partial ^{2}\varepsilon _{\mathrm{g}%
}/\partial \mu ^{2}\right) $, or the occurrence of second-order quantum
phase transition.

{Now we introduce two kinds of parameters that can identify
quantum phase transitional behavior. They are the analogue of pair order
parameter in a Hermitian superconducting system $\left\langle F\right\rangle 
$\ and the average of particle density $\left\langle n\right\rangle $. based
on the non-Hermitian version of Hellmann Feynman theorem \cite{HF} and the
translational symmetry of the ground state, we have the relations} 
\begin{equation}
\left\langle F\right\rangle =\frac{1}{2}\left\langle \overline{\text{G}}%
\right\vert (c_{2j+1}^{\dagger }c_{2j}^{\dagger }+c_{2j-1}c_{2j})\left\vert 
\text{G}\right\rangle =\frac{\partial \varepsilon _{\mathrm{g}}}{\partial
\Delta },  \label{APD}
\end{equation}%
and

\begin{equation}
\left\langle n\right\rangle =\left\langle \overline{\text{G}}\right\vert
n_{l}\left\vert \text{G}\right\rangle =1-\frac{\partial \varepsilon _{%
\mathrm{g}}}{\partial \mu },  \label{OP}
\end{equation}
with $l\in \left[ 1,2N\right] $\ and $j\in \left[ 1,N\right] $, where $%
\left\langle ...\right\rangle $\ denotes the expectation value for the
ground state in the framework of biorthogonal inner product. The choice of
biorthogonal inner product\ guarantees the restriction $\left\langle
n\right\rangle \eqslantless 1$, in comparison with the Dirac inner product.

It is clear that quantities $\left\langle F\right\rangle $\ and $%
\left\langle n\right\rangle $\ are discontinuous at the boundary. It is
natural to ask what happens at {quantum phase transition boundary%
}? To answer the question, numerical simulation is performed for the finite
system. $\left\langle F\right\rangle $\ and $\left\langle n\right\rangle $\
as functions of $\left( \Delta ,\mu \right) $\ are plotted in Fig. \ref{fig4}%
, which indicate that two quantities can be served as order parameters for
characterizing the quantum phase transitions only when they experience a
jump at the quantum phase boundary.

{We note that the phase boundary in Eq. (\ref{phase boundary}) is 
$\Delta $ dependent, while the white dashed lines in Fig. \ref{fig2} is $%
\Delta $ independent. In order to understand such a difference, we consider
a Hermitian Kitaev model with Hamiltonian $\mathcal{H}=H+H^{\dag }$. The
reason can be understood from the following analysis. Based on the Fourier
transformation, we have }

\begin{equation}
H=H_{k=0}+\sum_{k\in (0,\pi ]}H_{k},
\end{equation}%
where 
\begin{eqnarray}
&&H_{k=0}=2J\beta _{0}^{\dagger }\alpha _{0}+\mathrm{H.c.}+\Delta \beta
_{0}\alpha _{0}  \notag \\
&&+\Delta \beta _{0}^{\dagger }\alpha _{0}^{\dagger }+\mu \left( 2-2\beta
_{0}^{\dagger }\beta _{0}-2\alpha _{0}^{\dagger }\alpha _{0}\right) .
\end{eqnarray}%
This term is neglected in the above as usual for the safe of simplicity,
since the role of $H_{k=0}$\ can be obtained from the limit $k\rightarrow 0$%
. The boundary for $H$\ can be determined by $H_{k=0}$\ since the eigen
energy $\varepsilon _{k=0}=$ $\sqrt{4\mu ^{2}-\Delta ^{2}}-2J$, which is the
minimum of the excitations. From $\varepsilon _{k=0}=0$\ we then have $%
\Delta _{c}^{2}+4J_{c}^{2}-4\mu _{c}^{2}=0$. However, for $\mathcal{H}$\ the
pair term in $H_{k=0}+H_{k=0}^{\dag }$ is canceled, which results in the
conclusion that the boundary for $\mathcal{H}$\ is independent of $\Delta $.
In this sense, the stability of the boundary for $\mathcal{H}$\ comes from
the balance (or cancelation) of pair terms in $H$ and $H^{\dag }$,\
respectively.

\section{Topological phase}

\label{Topological phase}

{In this section, we investigate the quantum phase from
perspective of topology. }It is well known that the superconducting phase {%
in the Hermitian model} is topologically non-trivial, which is
characterized by winding number and demonstrated by edge modes in Majorana
lattice \cite{Kitaev}. In this section, we investigate this issue for the
present model. As mentioned above, the pair dispersion contains the term $%
r_{k}=$ $\sqrt{x_{k}^{2}+y_{k}^{2}}$, which has a zero point when the system
locates at the phase boundary. The previous work \cite{ZGPRL} implies that
there is a hidden ellipse with parameter equation $\left( x_{k},y_{k}\right) 
$ in the ground state. Next, we will extract the topological index from the
wave function of the pair ground state. 
\begin{figure*}[tbh]
\centering \includegraphics[width=1\textwidth]{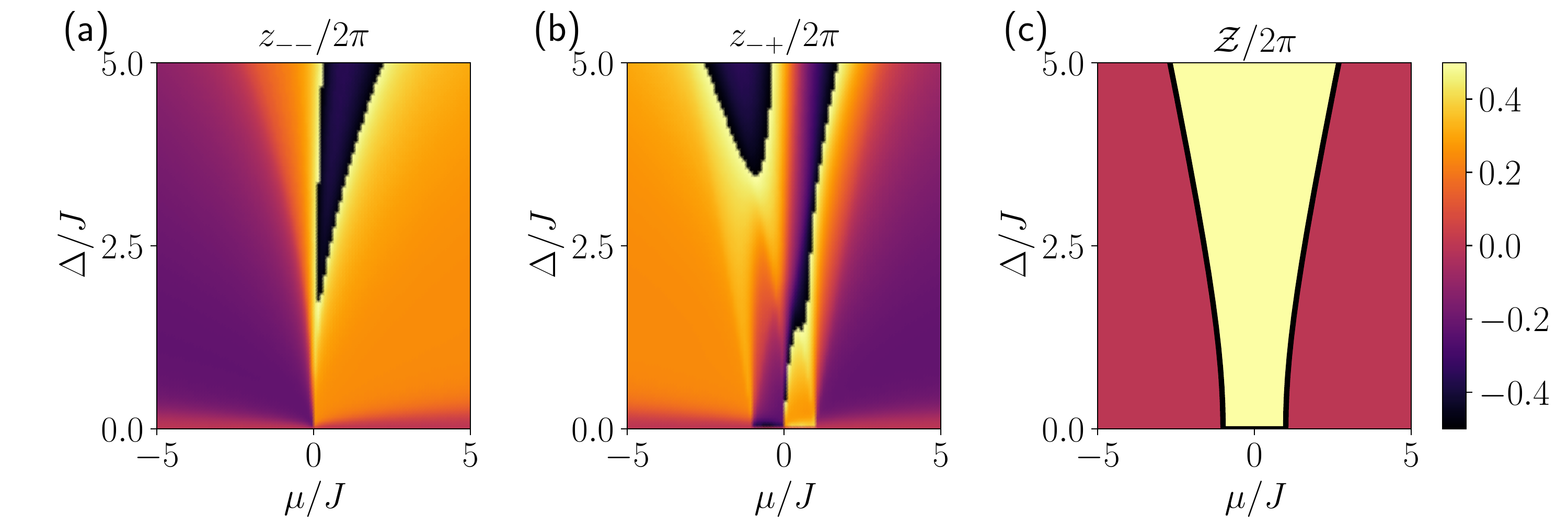}
\caption{2D contour color plots of (a) $z_{--}$, (b) $z_{-+}$, and (c) $%
\mathcal{Z}$, as functions of $\protect\mu $ and $\Delta $, {%
obtained from numerical results of Eqs.} (\protect\ref{Z_rs}) and (\protect
\ref{Z_sum}) on $N=51$ lattice. It can be seen that $z_{-+}$ and $z_{--}$\
are both irregular, while $\mathcal{Z}$\ is quantized and accords with the
phase boundary in (\protect\ref{phase boundary})\ indicated by the black
solid lines.}
\label{fig5}
\end{figure*}
To characterize the property of the energy band, we give the expression of
Zak phase%
\begin{equation}
z_{\rho \sigma }=\int_{0}^{2\pi }\left\langle \overline{\text{Vac}}%
\right\vert A_{\rho \sigma }^{k}\frac{\partial }{\partial k}\overline{A}%
_{\rho \sigma }^{k}\left\vert \mathrm{Vac}\right\rangle \mathrm{d}k.
\label{Z_rs}
\end{equation}%
$z_{\rho \sigma }$ is not an integer over $\pi $ in general case. In this
work, to characterize the property of the ground state, we study the Zak
phase of the lowest energy band of the pair spectrum. Our strategy is to
calculate the Berry flux of pair state $\left\vert \Psi _{k}\right\rangle =%
\overline{A}_{-+}^{k}\overline{A}_{--}^{k}\left\vert \text{Vac}\right\rangle 
$ 
\begin{equation}
\mathcal{Z}=\int_{0}^{2\pi }\mathcal{A}_{k}\mathrm{d}k,  \label{Z_sum}
\end{equation}%
where the Berry connection is given by%
\begin{equation}
\mathcal{A}_{k}=\left\langle \overline{\Psi _{k}}\right\vert \frac{\partial 
}{\partial k}\left\vert \Psi _{k}\right\rangle ,
\end{equation}%
with $\left\langle \overline{\Psi _{k}}\right\vert =\left\langle \overline{%
\text{Vac}}\right\vert A_{--}^{k}A_{-+}^{k}$. Notably, pair Zak phase $%
\mathcal{Z}$ can also be obtained via 
\begin{equation}
\mathcal{Z=}z_{-+}+z_{--}.
\end{equation}

In Fig. \ref{fig5} (c), the total Zak phase is numerically demonstrated in
parameter space.\textbf{\ }The solid black line $\Delta =2\sqrt{\mu
^{2}-J^{2}}$, indicates the boundary between the trivial and non-trivial
topological phase regions. It can be easily find that this line is also the
phase boundary in Fig. \ref{fig2}. As a comparison, $z_{--}$ and $z_{-+}$
are also plotted in Fig. \ref{fig5} (a) and (b), respectively. We notice
that the symmetry breaking has no effect on the ground state and $\mathcal{Z}
$, which accords with the edge states discussed in the following sections.

\begin{figure}[tbh]
\centering \includegraphics[width=0.45\textwidth]{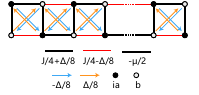} 
\caption{Geometry for the Majorana lattice described in Eq. (\protect\ref{MA}) with an open boundary. The system consists of two sublattices, $ia$ and $b$, indicated by solid and empty circles, respectively. The black solid lines,
the red solid lines and the double lines indicates the couplings between the
sublattices. The yellow and blue arrows are the unidirectional hoppings. The
eigenmodes near zero eigenvalue are plotted in Fig. \protect\ref{fig7}.} %
\label{fig6}
\end{figure}

\begin{figure}[tbh]
\centering\includegraphics[width=0.43\textwidth]{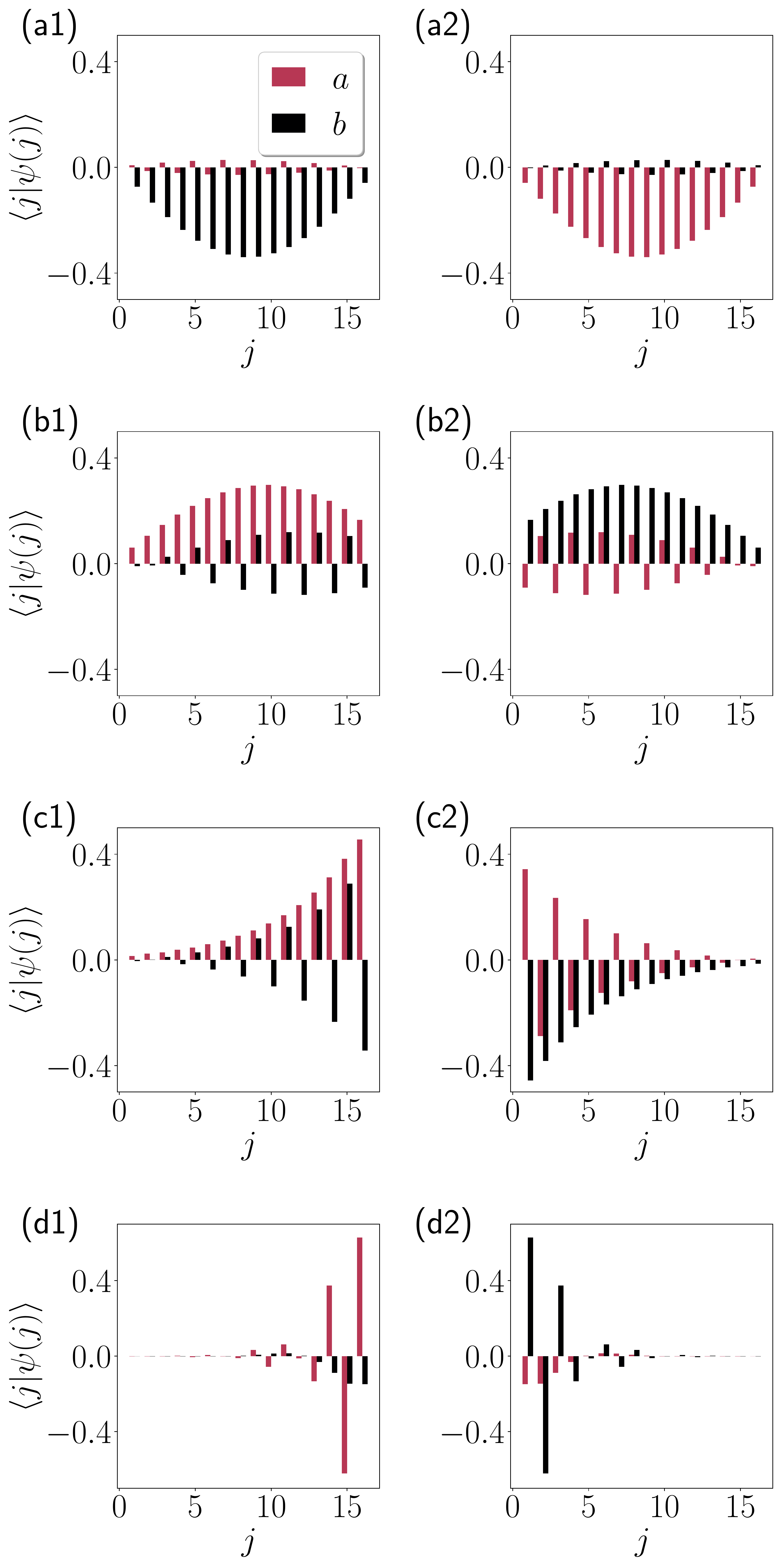} 
\caption{The amplitude distribution of eigenmodes of the Majorana lattices
in Fig. \protect\ref{fig6} near the zero eigenvalue obtained from exact
diagonalization on $N=8$ lattice at four different colored regions in Fig. 
\protect\ref{fig2}. (a) $\protect\mu = 3,\Delta = 1$, in blue region;
(b) $\protect\mu = 3,\Delta = 4$, in pink region; {\color{red}(c) $\protect\mu =
3,\Delta = 7$}, in green region and (d)$\protect\mu = -0.8,\Delta = 1$, in
red region. The color bar indicates the amplitude of wave function.
The profile of edge modes accords with the the expression in Eqs. (\protect
\ref{psil}) and (\protect\ref{psir}). It is shown that existence of nonzero
quantized Zak phase always associated with edge modes, demonstrating the
bulk edge correspondence.} \label{fig7}
\end{figure}

\section{Bulk-boundary correspondence}

\label{Bulk-boundary correspondence}

Let us now turn to the discussion on bulk-edge correspondence in such a
non-Hermitian system to investigate the influence from the non-Hermiticity.
Based on the above analysis, it turns out that the bulk system exhibits the
similar topological feature within the symmetry-unbroken regions. In
Hermitian systems, the existence of edge modes is intimately related to the
bulk topological quantum numbers, which is referred to as the bulk-edge
correspondence \cite{Thouless,Kane,Zhang,Lu}. We are interested in the
generalization of the bulk-edge correspondence to non-Hermitian systems.
Previous works have shown that when sufficiently weak non-Hermiticity is
introduced to topological insulator models, the edge modes can retain some
of their original characteristics \cite{Esaki,Hu}. For a non-Hermitian
Hamiltonian with full real spectrum, there is a Hermitian counterpart within
the symmetry-unbroken region \cite%
{Bender,Ali1,Bender2,Jones,Jin1,Jin2,Jin3,Ali6,Ali7,Ali8}. Then the bulk-edge
correspondence can be established based on the Hermitian counterpart \cite%
{LCPRB}. Next, we will show that the topological phase still exists even in
the symmetric broken region for the present Hamiltonian, with the existence
of edge modes.

Considering the spinless fermion system with an open boundary condition, the
Hamiltonians read%
\begin{eqnarray}
&&H_{\mathrm{CH}}=H-M, \\
&&M=\Delta c_{2N}^{\dagger }c_{1}^{\dagger }+Jc_{2N}^{\dagger
}c_{1}+Jc_{1}^{\dagger }c_{2N}.  \label{H_K}
\end{eqnarray}%
We introduce Majorana fermion operators%
\begin{equation}
a_{j}=c_{j}^{\dagger }+c_{j},b_{j}=-i\left( c_{j}^{\dagger }-c_{j}\right) ,
\label{ab}
\end{equation}%
which satisfy the relations%
\begin{eqnarray}
\left\{ a_{j},a_{j^{\prime }}\right\} &=&2\delta _{j,j^{\prime }},\left\{
b_{j},b_{j^{\prime }}\right\} =2\delta _{j,j^{\prime }}, \\
\left\{ a_{j},b_{j^{\prime }}\right\} &=&0.
\end{eqnarray}%
The inverse transformation is 
\begin{equation}
c_{j}^{\dagger }=\frac{1}{2}\left( a_{j}+ib_{j}\right) ,c_{j}=\frac{1}{2}%
\left( a_{j}-ib_{j}\right) .
\end{equation}%
{the Majorana representation of the Hamiltonian has the form }

\begin{eqnarray}
H_{\mathrm{CH}} &=&\sum_{j=1}^{N}[i(\frac{J}{4}+\frac{\Delta }{8})\left(
-a_{2j-1}b_{2j}-a_{2j}b_{2j+1}\right)  \notag \\
&&+i(\frac{J}{4}-\frac{\Delta }{8})\left(
b_{2j-1}a_{2j}+b_{2j}a_{2j+1}\right) ]+\mathrm{H.c.}  \notag \\
&&-\frac{\Delta }{4}\left(
a_{2j}a_{2j-1}+a_{2j}a_{2j+1}+b_{2j+1}b_{2j}+b_{2j-1}b_{2j}\right)  \notag \\
&&-\frac{\mu }{2}\sum_{j=1}^{2N}\left( ib_{j}a_{j}+\mathrm{H.c.}\right) 
\mathrm{.}  \label{H_M}
\end{eqnarray}%
{Based on the identities }

\begin{eqnarray}
&&c_{l}^{\dag }c_{l+1}+\text{\textrm{H.c.}}=ib_{j}a_{j+1}  \notag \\
&=&\frac{1}{2}\left( b_{j},-ia_{j+1}\right) \left( 
\begin{array}{cc}
0 & 1 \\ 
1 & 0%
\end{array}%
\right) \left( 
\begin{array}{c}
b_{j} \\ 
ia_{j+1}%
\end{array}%
\right) , \\
n_{l} &=&ib_{j}a_{j}=\frac{1}{2}\left( -ia_{j},b_{j}\right) \left( 
\begin{array}{cc}
0 & 1 \\ 
1 & 0%
\end{array}%
\right) \left( 
\begin{array}{c}
ia_{j} \\ 
b_{j}%
\end{array}%
\right) ,
\end{eqnarray}%
and%
\begin{eqnarray}
c_{2j+1}^{\dagger }c_{2j}^{\dagger } &=&\frac{1}{8}\varphi _{1}^{\dag
}\left( 
\begin{array}{cccc}
0 & 0 & -1 & 1 \\ 
0 & 0 & -1 & 1 \\ 
1 & -1 & 0 & 0 \\ 
1 & -1 & 0 & 0%
\end{array}%
\right) \varphi _{1}, \\
c_{2j-1}c_{2j} &=&\frac{1}{8}\varphi _{2}^{\dag }\left( 
\begin{array}{cccc}
0 & 0 & 1 & 1 \\ 
0 & 0 & -1 & -1 \\ 
-1 & -1 & 0 & 0 \\ 
1 & 1 & 0 & 0%
\end{array}%
\right) \varphi _{2},
\end{eqnarray}%
where $\varphi _{1}^{\dag }=$ $\left( -ia_{2j},b_{2j},\text{ }%
-ia_{2j+1},b_{2j+1}\right) $\ and $\varphi _{2}^{\dag }=$ $\left(
-ia_{2j-1},b_{2j-1},\text{ }-ia_{2j},b_{2j}\right) $, we write down the
Hamiltonian in the basis $\varphi ^{\dagger }=(-ia_{1},$ $b_{1},$ $-ia_{2},$ 
$b_{2},$ $-ia_{3},$ $b_{3},$ $...)$ and see that%
\begin{equation}
H_{\mathrm{CH}}=\varphi ^{\dagger }h_{\mathrm{CH}}\varphi ,
\end{equation}%
where $h_{\mathrm{CH}}$\ represents a $4N\times 4N$ matrix. Here matrix $h_{%
\mathrm{CH}}$\ can be explicitly written as

\begin{eqnarray}
&&h_{\text{\textrm{CH}}}=(\frac{J}{4}+\frac{\Delta }{8})\sum_{j=1}^{N}\left(
\left\vert 2j-1\right\rangle _{AB}\left\langle 2j\right\vert +\left\vert
2j\right\rangle _{AB}\left\langle 2j+1\right\vert \right)  \notag \\
&&+(\frac{J}{4}-\frac{\Delta }{8})\sum_{j=1}^{N}\left( \left\vert
2j-1\right\rangle _{BA}\left\langle 2j\right\vert +\left\vert
2j\right\rangle _{BA}\left\langle 2j+1\right\vert \right)  \notag \\
&&-\frac{\mu }{2}\sum_{j=1}^{2N}\left\vert j\right\rangle _{AB}\left\langle
j\right\vert +\mathrm{H.c.}  \notag \\
&&-\frac{\Delta }{8}\sum_{j=1}^{N}\left( \left\vert 2j\right\rangle
_{AA}\left\langle 2j-1\right\vert +\left\vert 2j\right\rangle
_{AA}\left\langle 2j+1\right\vert -\mathrm{H.c.}\right)  \notag \\
&&-\frac{\Delta }{8}\sum_{j=1}^{N}\left( \left\vert 2j-1\right\rangle
_{BB}\left\langle 2j\right\vert +\left\vert 2j+1\right\rangle
_{BB}\left\langle 2j\right\vert -\mathrm{H.c.}\right) \mathrm{,}  \label{MA}
\end{eqnarray}%
where basis $\left\{ \left\vert l\right\rangle _{A},\left\vert
l\right\rangle _{B},l\in \left[ 1,2N\right] \right\} \ $is an orthonormal
complete set, $_{A}\langle l\left\vert l^{\prime }\right\rangle _{B}=\delta
_{ll^{\prime }}\delta _{AB}$, which accords with $\varphi $. Schematic
illustration for structure of $h_{\mathrm{CH}}$\ is described in Fig. \ref%
{fig6} by different types of nodal lines.

The edge states at the limitation of infinity large size system $\left(
N\longrightarrow \infty \right) $ are analytically obtained. For the case
with $\Delta >2\sqrt{\mu ^{2}-J^{2}}$, the edge states of $h_{\mathrm{CH}}$\
are in the form of%
\begin{eqnarray}
\left\vert \psi _{\mathrm{L}}\right\rangle &=&\frac{1}{\sqrt{\Omega }}%
\sum_{j=1}^{2N}[\left( \gamma _{1}\right) ^{j}-\left( \gamma _{2}\right)
^{j}]\widetilde{\left\vert j_{1}\right\rangle },  \label{psil} \\
\left\vert \psi _{\mathrm{R}}\right\rangle &=&\frac{1}{\sqrt{\Omega }}%
\sum_{j=1}^{2N}[\left( \gamma _{1}\right) ^{2N+1-j}-\left( \gamma
_{2}\right) ^{2N+1-j}]\widetilde{\left\vert j_{2}\right\rangle },
\label{psir}
\end{eqnarray}%
where the kets are%
\begin{eqnarray}
\widetilde{\left\vert j_{1}\right\rangle } &=&\left\vert j\right\rangle
_{A}+\left( -1\right) ^{j}\beta \left\vert j\right\rangle _{B} \\
\widetilde{\left\vert j_{2}\right\rangle } &=&\left\vert j\right\rangle
_{B}+\left( -1\right) ^{2N+1-j}\beta \left\vert j\right\rangle _{A}
\end{eqnarray}%
with the coefficients%
\begin{equation}
\gamma _{l}=\frac{2\mu +(-1)^{l}2\sqrt{\mu ^{2}-J^{2}}}{\sqrt{\Delta
^{2}+4J^{2}}+\Delta },
\end{equation}%
and%
\begin{equation}
\beta =\sqrt{1+\left( 2J/\Delta \right) ^{2}}+2J/\Delta ,
\end{equation}%
and $\Omega $ is the normalization factor. The derivations of the analysis
form of the edge states is presented in Appendix \ref{AppendixB}. We plot
the distribution of particle amplitude $p\left( j\right) $ of $\left\vert
\psi _{\mathrm{L}}\right\rangle $ and $\left\vert \psi _{\mathrm{R}%
}\right\rangle $\ in Fig. \ref{fig7}. In contrast there is no zero mode for $%
\Delta <2\sqrt{\mu ^{2}-J^{2}}$, which indicates the coexistence of zero
modes and nonzero Zak phases, demonstrating the bulk-edge correspondence.

\begin{figure*}[tbh]
\centering \includegraphics[width=0.8\textwidth]{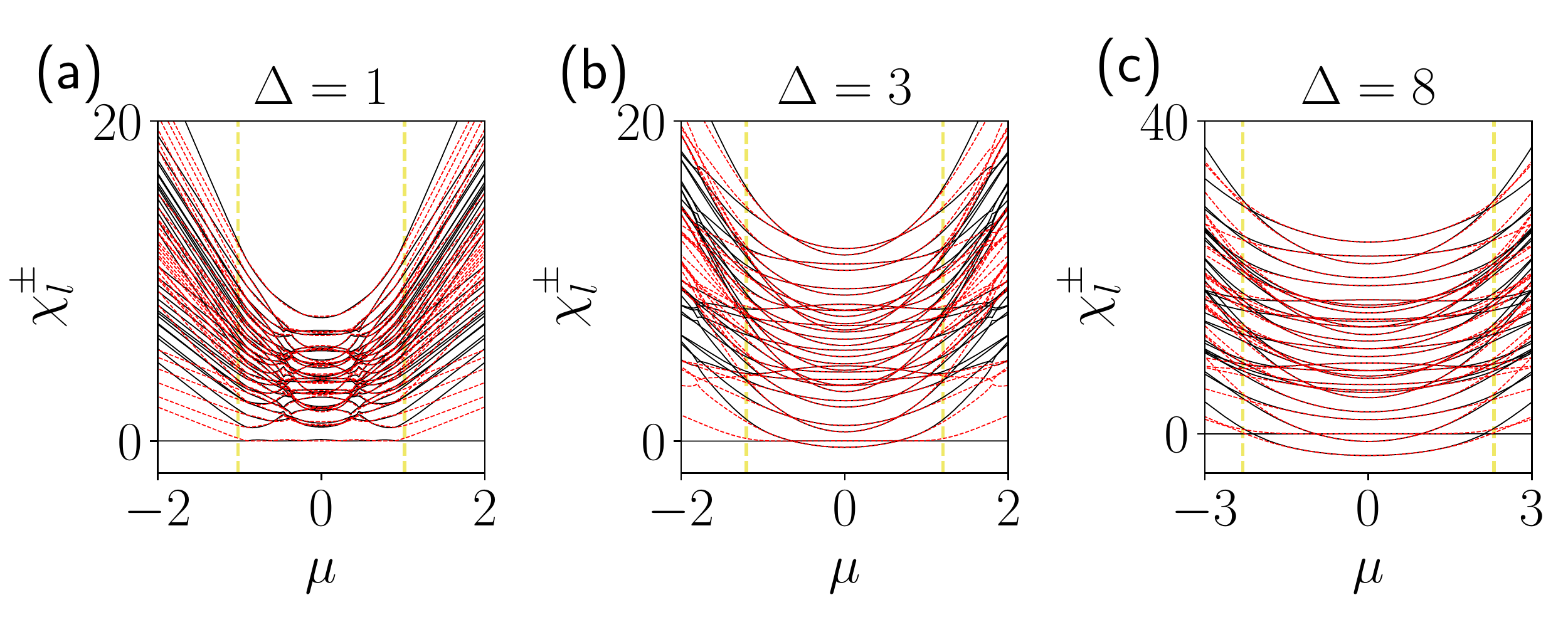}
\caption{The plot of $\protect\chi _{l}^{\pm }$ as a function of $\protect%
\mu $, obtained through exact diagonalization. System parameter: $N=6$ and $%
J=1$. Red dashed lines represent the values of $\left\{ \protect\chi %
_{l}^{+}\right\} $ and the black lines represent the values of $\left\{ 
\protect\chi _{l}^{-}\right\} $. Notably, all energy levels become twofold
degeneracy simultaneously at one point which is marked by the yellow dashed
lines, protected by the symmetry of the operators ${\overline{D},D.}$ {\ 
The degeneracy point gradually approaches the topological
non-trivial boundary as $N$ increases.}}
\label{fig8}
\end{figure*}

\section{Kramer-like degeneracy}

\label{Kramer-like degeneracy}

In this section, we further elucidate the implication of the Majorana zero
modes on the feature of the system and how it provides an evident signature
of the phase diagram regardless of the appearance of the complex energy
levels.

The parity and time reversal symmetries of $H$ always hold for finite size
ring or chain. Inspired from the previous works \cite{ZKLPRL,ZKLPRB}, the
present Hamiltonian in thermodynamic limit also possesses a conditional
symmetry in the topological phase, although it is non-Hermitian system. It
is due to the existence of zero modes in the Majorana lattice. Based on the
exact expression of $\left\vert \psi _{\mathrm{L,R}}\right\rangle $\textbf{\ 
}and\textbf{\ }$\overline{\left\vert \psi _{\mathrm{L,R}}\right\rangle }$,\
one can construct a set of nonlocal operators%
\begin{eqnarray}
D &=&\frac{i}{2\sqrt{\Omega }}\sum_{j=1}^{2N}[\left( \gamma _{1}\right)
^{j}-\left( \gamma _{2}\right) ^{j}][\left( -1\right) ^{j}\beta
d_{1}^{\dagger }+d_{2}], \\
\overline{D} &=&\frac{i}{2\sqrt{\Omega }}\sum_{j=1}^{2N}[\left( \gamma
_{1}\right) ^{j}-\left( \gamma _{2}\right) ^{j}][\left( -1\right) ^{j}\beta
d_{1}-d_{2}^{\dagger }],
\end{eqnarray}%
with%
\begin{eqnarray}
d_{1} &=&c_{2N+1-j}^{\dagger }-c_{j}^{\dagger }+c_{2N+1-j}+c_{j}, \\
d_{2} &=&c_{2N+1-j}-c_{j}-c_{2N+1-j}^{\dagger }-c_{j}^{\dagger },
\end{eqnarray}%
which satisfy the commutation relations%
\begin{equation}
\left[ D,H_{\mathrm{CH}}\right] =\left[ \overline{D},H_{\mathrm{CH}}\right]
=0,  \label{DH}
\end{equation}%
and%
\begin{equation}
\left\{ D,\overline{D}\right\} =1,\overline{D}^{2}=D^{2}=0.
\end{equation}%
Considering the set of complete eigenstates $\left\{ \left\vert \psi
_{n}^{+}\right\rangle ,\left\vert \psi _{n}^{-}\right\rangle \right\} $ of $%
H $ with eigen energy $\varepsilon _{n}^{\pm }$, $H\left\vert \psi _{n}^{\pm
}\right\rangle =\varepsilon _{n}^{\pm }\left\vert \psi _{n}^{\pm
}\right\rangle $, where 
\begin{equation}
\Pi \left\vert \psi _{n}^{\pm }\right\rangle =\pm \left\vert \psi _{n}^{\pm
}\right\rangle ,
\end{equation}%
we have the relations%
\begin{eqnarray}
D\left\vert \psi _{n}^{+}\right\rangle &=&\left\vert \psi
_{n}^{-}\right\rangle ,\overline{D}\left\vert \psi _{n}^{-}\right\rangle
=\left\vert \psi _{n}^{+}\right\rangle , \\
\overline{D}\left\vert \psi _{n}^{+}\right\rangle &=&D\left\vert \psi
_{n}^{-}\right\rangle =0,
\end{eqnarray}%
which guarantee the existence of eigenstates degeneracy $\varepsilon
_{n}^{+}=\varepsilon _{n}^{-}=\varepsilon _{n}$, referred to as Kramer-like
degeneracy \cite{ZKLPRL,ZKLPRB}. Note that here $\varepsilon _{n}$\ can be
complex energy.

Let us now examine the performance of our finding in finite size system. To
demonstrate this point, numerical results is presented. The spectrum of
Kitaev Hamiltonian $H_{\mathrm{CH}}$ on finite chain can be obtained
numerically. We reorder the complex energy level $\left\{ \varepsilon
_{n}^{\pm }\right\} $\ by the value of $\mathcal{E}_{l}^{\pm }=\mathrm{Re}\
\varepsilon _{n}^{\pm }+\mathrm{Im}\ \varepsilon _{n}^{\pm }$, and calculate
the mode of the relative value%
\begin{equation}
\mathcal{\chi }_{l}^{\pm }=\left\vert \mathcal{E}_{l}^{\pm }-\mathcal{E}%
_{1}\right\vert
\end{equation}%
where $\mathcal{E}_{1}$ means the smallest value in $\left\{ \mathcal{E}%
_{n}^{\pm }\right\} $. {Fig. \ref{fig8} (a), (b) and (c) plot $%
\mathcal{\chi }_{l}^{\pm }(\mu ,\Delta)$ for $\Delta=J$, $3J$ and $8J$, as%
\textbf{\ }functions of $\mu $.} As expected, the complete spectrum $%
\mathcal{\chi }_{l}^{\pm }$\ becomes two-fold degeneracy approximately
within the non-trivial regions. {The $\mu$-parameter regime of
degeneracy that been illustrated by yellow dashed lines in Fig. \ref{fig8}
widens with $\Delta$.}

\section{Summary}

\label{sec_summary}

In summary, we have analyzed a one-dimensional non-Hermitian Kitaev model
with locally imbalanced but globally balanced pair creation and
annihilation, which is introduced by staggered non-Hermitian pairing terms.
It exhibits the similar topological features compared with the original
Hermitian one, but modifies the phase boundary. In parallel, a non-trivial
topological phase is always associated with edge states in the open chain,
in which the Zak phase is defined in the terms of biorthonormal inner
product for the BCS-like pair wave function. Correspondingly, the edge modes
are obtained with the aid of the Majorana transformation, resulting in the
Kramer-like degeneracy of both the real and complex levels as a signature of
the topologically non-trivial phase. These results indicate that the
staggered non-Hermitian pairing terms still create superconducting phases
and their topological feature is immune to the non-Hermitian effect. This
study provides the insight into the topological phase emerged from the
interplay between spatially separated pairing processes. {
Experimentally, the Majorana can be realized in photonic system and the edge
modes can be detected \cite{SW}. Recently, it has been shown that the phase
diagram of a Hermitian Kitaev model can be demonstrated in the dynamic
process \cite{SYBPRB}. This will be an interesting topic for our future work
to investigate the dynamics of the present model.}

\section{Appendix}

\appendix

In this Appendix, the derivations of Bogoliubov modes in Eq. (\ref{1E}),
edge modes in Eqs. (\ref{psil})\ and (\ref{psir}) are presented.

\subsection{Bogoliubov modes}

\label{AppendixA} \setcounter{equation}{0} \renewcommand{\theequation}{A%
\arabic{equation}} \renewcommand{\thesubsection}{\arabic{subsection}}
Consider the matrix%
\begin{equation}
h_{k}=\left( 
\begin{array}{cccc}
-2\mu & \gamma _{k} & 0 & -\Delta _{\mathrm{o}}e^{ik} \\ 
\gamma _{k}^{\ast } & -2\mu & \Delta _{\mathrm{o}}e^{-ik} & 0 \\ 
0 & -\Delta _{\mathrm{e}} & 2\mu & -\gamma _{k} \\ 
\Delta _{\mathrm{e}} & 0 & -\gamma _{k}^{\ast } & 2\mu%
\end{array}%
\right) ,
\end{equation}%
which can be written as the form 
\begin{equation}
h_{k}=\left( 
\begin{array}{cccc}
-a & b & 0 & -c \\ 
b^{\ast } & -2\mu & c^{\ast } & 0 \\ 
0 & -d & 2\mu & -b_{k} \\ 
d & 0 & -b_{k}^{\ast } & a%
\end{array}%
\right) ,
\end{equation}%
by the substitutions%
\begin{equation}
a=2\mu ,b=\gamma _{k},c=\Delta _{\mathrm{o}}e^{ik},d=\Delta _{\mathrm{e}}.
\end{equation}%
Here we remain using the parameters $\Delta _{\mathrm{o}}$\ and $\Delta _{%
\mathrm{e}}$ rather than $\Delta $. We will show that the solution only
depends on $\Delta _{\mathrm{o}}\Delta _{\mathrm{e}}$. Taking a similarity
transformation the matrix $h_{k}$\ is diagonalized as%
\begin{equation}
V^{k}h_{k}\overline{V}^{k}=\left( 
\begin{array}{cccc}
\varepsilon _{++}^{k} & 0 & 0 & 0 \\ 
0 & \varepsilon _{+-}^{k} & 0 & 0 \\ 
0 & 0 & \varepsilon _{--}^{k} & 0 \\ 
0 & 0 & 0 & \varepsilon _{-+}^{k}%
\end{array}%
\right) ,
\end{equation}%
where the eigenvalue is%
\begin{equation}
\varepsilon _{\rho \sigma }^{k}=\rho \sqrt{\Lambda _{k}+\sigma \sqrt{\Omega
_{k}}},
\end{equation}%
with $\rho ,\sigma =\pm $\textbf{\ }and\textbf{\ }$\overline{V}^{k}=\left(
V^{k}\right) ^{-1}$\textbf{.} Here the parameters are

{
\begin{eqnarray}
\Lambda _{k} &=&a^{2}+\left\vert b\right\vert ^{2}-\frac{\left( cd+c^{\ast
}d\right) }{2}  \notag \\
&=&4\mu ^{2}-\Delta _{\mathrm{o}}\Delta _{\mathrm{e}}\cos k+4J^{2}\cos ^{2}%
\frac{k}{2},
\end{eqnarray}%
} and%
\begin{eqnarray}
\Omega _{k} &=&-cd\left( b^{\ast }\right) ^{2}+4a^{2}\left\vert b\right\vert
^{2}-cd\left\vert b\right\vert ^{2}-dc^{\ast }\left\vert b\right\vert ^{2} 
\notag \\
&&-c^{\ast }db^{2}-\frac{d^{2}\left\vert c\right\vert ^{2}}{2}+\frac{%
c^{2}d^{2}+c^{\ast 2}d^{2}}{4}  \notag \\
&=&64\mu ^{2}J^{2}\cos ^{2}\frac{k}{2}-\left( \Delta _{\mathrm{o}}\Delta _{%
\mathrm{e}}\right) ^{2}\sin ^{2}k  \notag \\
&&-16J^{2}\Delta _{\mathrm{o}}\Delta _{\mathrm{e}}\cos ^{4}\frac{k}{2},
\end{eqnarray}%
which only depends on $\Delta _{\mathrm{o}}\Delta _{\mathrm{e}}$, and then
can be replaced by $\Delta =\sqrt{\Delta _{\mathrm{o}}\Delta _{\mathrm{e}}}$
for positive $\Delta _{\mathrm{o}}$\ and $\Delta _{\mathrm{e}}$. The
explicit form of $V^{k}$ and $\overline{V}^{k}$ can be obtained from the
eigenvector $\Psi _{\rho \sigma }$\ of eigenvalue $\varepsilon _{\rho \sigma
}^{k}$,%
\begin{eqnarray}
&&\Psi _{\rho \sigma }= \\
&&\left( 
\begin{array}{c}
8a\left\vert b\right\vert ^{2}+2(dc^{\ast }-cd+2\rho \sqrt{\Omega _{k}}%
)(a-\varepsilon _{\rho \sigma }) \\ 
2\left[ 2bdc^{\ast }+b^{\ast }(cd+dc^{\ast }-4a^{2}-2\rho \sqrt{\Omega _{k}}%
+4a\varepsilon _{\rho \sigma })\right] \\ 
4d\left[ ab-ab^{\ast }+(b+b^{\ast })\varepsilon _{\rho \sigma }\right] \\ 
2d\left[ cd-2\left\vert b\right\vert ^{2}-2\left( b^{\ast }\right)
^{2}-dc^{\ast }-2\rho \sqrt{\Omega _{k}}\right]%
\end{array}%
\right) ,  \notag
\end{eqnarray}%
or%
\begin{equation}
\left( 
\begin{array}{c}
64\mu J^{2}\cos ^{2}\frac{k}{2}+(4\mu -2\varepsilon _{\rho \sigma })(\rho 
\sqrt{\Omega _{k}}-i2\Delta ^{2}\sin k) \\ 
2J\gamma _{k}\left( -16\mu ^{2}-\rho \sqrt{\Omega _{k}}+8\mu \varepsilon
_{\rho \sigma }+8\Delta ^{2}\cos ^{2}\frac{k}{2}\right) \\ 
16\Delta J\left( \varepsilon _{\rho \sigma }\cos ^{2}\frac{k}{2}+i\mu \sin
k\right) \\ 
2\Delta \left( -\rho \sqrt{\Omega _{k}}-8J^{2}\gamma _{k}\cos ^{2}\frac{k}{2}%
+i2\Delta ^{2}\sin k\right)%
\end{array}%
\right) .
\end{equation}

\subsection{The derivations of the edge states}

\label{AppendixB} \setcounter{equation}{0} \renewcommand{\theequation}{B%
\arabic{equation}} \renewcommand{\thesubsection}{\arabic{subsection}}

The Bethe ansatz wave function of the edge mode localized at left side $%
\left\vert \psi _{\mathrm{L}}\right\rangle $ for the matrix $h_{\text{%
\textrm{CH}}}$\ has the form 
\begin{eqnarray}
&&\left\vert \psi _{\mathrm{L}}\right\rangle =\frac{1}{\sqrt{\Omega }}%
\sum_{j=1}^{2N}\sum_{l=1,2}\alpha _{l}\left( \gamma _{l}\right)
^{2j}(\left\vert 2j-1\right\rangle _{A}  \notag \\
&&+\beta \left\vert 2j-1\right\rangle _{B}+\gamma _{l}\left\vert
2j\right\rangle _{A}-\beta \gamma _{l}\left\vert 2j\right\rangle _{B}),
\end{eqnarray}%
where $\Omega $ is the normalization factor. The Schr\H{o}dinger equation 
\begin{equation}
h_{\text{\textrm{CH}}}\left\vert \psi _{\mathrm{L}}\right\rangle =0,
\end{equation}%
gives that $\gamma _{1}$ and $\gamma _{2}$ satisfy the same equations%
\begin{eqnarray}
\frac{4\mu \gamma _{l}}{1+\left( \gamma _{l}\right) ^{2}} &=&\Delta \lbrack 
\frac{1-\left( \gamma _{l}\right) ^{2}}{1+\left( \gamma _{l}\right) ^{2}}%
+\beta ]+2J, \\
\frac{4\mu \gamma _{l}}{1+\left( \gamma _{l}\right) ^{2}} &=&\Delta \lbrack 
\frac{1-\left( \gamma _{l}\right) ^{2}}{1+\left( \gamma _{l}\right) ^{2}}+%
\frac{1}{\beta }]-2J.
\end{eqnarray}%
After a straightforward algebra, we have the non-trivial solutions

\begin{equation}
\gamma _{l}=\frac{2\mu +(-1)^{l}2\sqrt{\mu ^{2}-J^{2}}}{\sqrt{\Delta
^{2}+4J^{2}}+\Delta },
\end{equation}%
and%
\begin{equation}
\beta =\sqrt{1+\left( 2J/\Delta \right) ^{2}}+2J/\Delta .
\end{equation}%
The ratio of $\alpha _{1}$ and $\alpha _{2}$ is 
\begin{equation}
\alpha _{1}\gamma _{1}+\alpha _{2}\gamma _{2}=0,
\end{equation}%
obtained by the boundary condition at $j=1$. Similarly, the wave function of
the edge mode localized at right side $\left\vert \psi _{\mathrm{R}%
}\right\rangle $ can also be obtained. In summary we have%
\begin{eqnarray}
\left\vert \psi _{\mathrm{L}}\right\rangle &=&\frac{1}{\sqrt{\Omega }}%
\sum_{j=1}^{2N}\left[ \left( \gamma _{1}\right) ^{j}-\left( \gamma
_{2}\right) ^{j}\right] \widetilde{\left\vert j_{1}\right\rangle }, \\
\left\vert \psi _{\mathrm{R}}\right\rangle &=&\frac{1}{\sqrt{\Omega }}%
\sum_{j=1}^{2N}[\left( \gamma _{1}\right) ^{2N+1-j}  \notag \\
&&-\left( \gamma _{2}\right) ^{2N+1-j}]\widetilde{\left\vert
j_{2}\right\rangle },
\end{eqnarray}%
where the kets%
\begin{eqnarray}
\widetilde{\left\vert j_{1}\right\rangle } &=&\left\vert j\right\rangle
_{A}+\left( -1\right) ^{j}\beta \left\vert j\right\rangle _{B}, \\
\widetilde{\left\vert j_{2}\right\rangle } &=&\left\vert j\right\rangle
_{B}+\left( -1\right) ^{2N+1-j}\beta \left\vert j\right\rangle _{A}.
\end{eqnarray}%
Two states $\left\vert \psi _{\mathrm{L}}\right\rangle $\textbf{\ }and%
\textbf{\ }$\left\vert \psi _{\mathrm{R}}\right\rangle $\ are localized
states in the non-trivial regions due to the following facts. (i) In the
case with $J<$ $\left\vert \mu \right\vert <$ $\sqrt{\Delta ^{2}/4+J^{2}}$, $%
\gamma _{1}$\ and $\gamma _{2}$\ are real and it is easy to check that $%
\left\vert \gamma _{1}\right\vert <1$\ and $\left\vert \gamma
_{2}\right\vert <1$. (ii) In the case with\textbf{\ }$\left\vert \mu
\right\vert <J$, $\gamma _{1}$\ and $\gamma _{2}$\ become complex number and
have the form%
\begin{equation}
\gamma _{1}=\left( \gamma _{2}\right) ^{\ast },
\end{equation}%
with\textbf{\ }%
\begin{equation}
\left\vert \gamma _{1}\right\vert =\left\vert \gamma _{2}\right\vert =\frac{%
2J}{\sqrt{\Delta ^{2}+4J^{2}}+\Delta }<1,
\end{equation}%
and%
\begin{equation}
\arg \gamma _{2}=-\arg \gamma _{1}=\tan ^{-1}\frac{\sqrt{\mu ^{2}-J^{2}}}{%
\mu }.
\end{equation}%
It can be checked that 
\begin{equation}
\left( \gamma _{1}\right) ^{j}-\left( \gamma _{2}\right) ^{j}=i2\left\vert
\gamma _{1}\right\vert ^{j}\sin \left( j\arg \gamma _{2}\right) .
\end{equation}%
Then we conclude that $\left\vert \psi _{\mathrm{L}}\right\rangle $\ and $%
\left\vert \psi _{\mathrm{R}}\right\rangle $\ are the edge states. Notably,
the existence of the edge modes is independent of the phase diagram arising
from the breaking of $\mathcal{T}$\ symmetry. The eigenvalues of edge modes
are always zero without imaginary part. Similarly\textbf{, }the zero modes
of the matrix $h_{\mathrm{CH}}^{\dagger }$\ can be obtained as form%
\begin{eqnarray}
\overline{\left\vert \psi _{\mathrm{L}}\right\rangle } &=&\frac{1}{\sqrt{%
\Omega }}\sum_{j=1}^{2N}[\left( \gamma _{1}\right) ^{j}-\left( \gamma
_{2}\right) ^{j}]\overline{\left\vert j_{1}\right\rangle }, \\
\overline{\left\vert \psi _{\mathrm{R}}\right\rangle } &=&\frac{1}{\sqrt{%
\Omega }}\sum_{j=1}^{2N}[\left( \gamma _{1}\right) ^{2N+1-j}  \notag \\
&&-\left( \gamma _{2}\right) ^{2N+1-j}]\overline{\left\vert
j_{2}\right\rangle },
\end{eqnarray}%
where%
\begin{eqnarray}
\overline{\left\vert j_{1}\right\rangle } &=&\left\vert j\right\rangle
_{A}-\left( -1\right) ^{j}\beta \left\vert j\right\rangle _{B}, \\
\overline{\left\vert j_{2}\right\rangle } &=&\left\vert j\right\rangle
_{B}-\left( -1\right) ^{2N+1-j}\beta \left\vert j\right\rangle _{A}.
\end{eqnarray}

\acknowledgments We acknowledge the support of the National Natural Science
Foundation of China (Grant No. 11874225).


\begin{thebibliography}{99}
\bibitem{DAMQ} {D. A. McQuarrie, Quantum Chemistry. University
Science Books, Mill Valey, CA, 1983.}

\bibitem{Bender} C. M. Bender and S. Boettcher, Real spectra in
non-Hermitian Hamiltonians having $\mathcal{PT}$ symmetry, Phys. Rev. Lett. 
\textbf{80,} 5243 (1998).

\bibitem{Bender1} C. M. Bender, D. C. Brody, and H. F. Jones, Complex
extension of quantum mechanics, Phys. Rev. Lett. \textbf{89,} 270401 (2002).

\bibitem{Ali1} A. Mostafazadeh, Pseudo-Hermiticity versus $\mathcal{PT}$%
-symmetry: Equivalence of pseudo-Hermiticity and the presence of antilinear
symmetries, J. Math. Phys. \textbf{43,} 205 (2002).

\bibitem{Ali2} A. Mostafazadeh, Pseudo-Hermiticity versus $\mathcal{PT}$%
-symmetry II: Equivalence of pseudo-Hermiticity and the presence of
antilinear symmetries, J. Math. Phys. \textbf{43,} 2814 (2002).

\bibitem{Ali3} A. Mostafazadeh, Pseudo-Hermiticity versus $\mathcal{PT}$%
-symmetry III: Equivalence of pseudo-Hermiticity and the presence of
antilinear symmetries, J. Math. Phys. \textbf{43,} 3944 (2002). 

\bibitem{Ali4} A. Mostafazadeh, Pseudo-supersymmetric quantum mechanics and
isospectral pseudo-Hermitian Hamiltonians, Nucl. Phys. B \textbf{640,} 419
(2002).

\bibitem{Ali5} A. Mostafazadeh, Pseudo-Hermiticity and generalized $\mathcal{PT}$- and
$\mathcal{CPT}$-symmetries, J. Math. Phys. \textbf{44,} 974 (2003).

\bibitem{Jin1} L. Jin and Z. Song, Solutions of $\mathcal{PT}$-symmetric
tight-binding chain and its equivalent Hermitian counterpart, Phys. Rev. A 
\textbf{80,} 052107 (2009).

\bibitem{Jin2} L. Jin and Z. Song, Physics counterpart of the $\mathcal{PT}$
non-Hermitian tight-binding chain, Phys. Rev. A \textbf{81,} 032109 (2010). 

\bibitem{Jin3} L. Jin and Z. Song, A physical interpretation for the
non-Hermitian Hamiltonian, J. Phys. A: Math. Theor. \textbf{44,} 375304
(2011).

\bibitem{Kitaev} A. Y. Kitaev, Unpaired Majorana fermions in quantum wires,
Phys. Usp. \textbf{44,} 131 (2001).

\bibitem{Sarma} C. Nayak, S. H. Simon, A. Stern, M. Freedman, and S. D.
Sarma, Non-Abelian anyons and topological quantum computation, Rev. Mod.
Phys. \textbf{80,} 1083 (2008).

\bibitem{Stern} A. Stern, Non-Abelian states of matter, Nature (London) 
\textbf{464,} 187 (2010).

\bibitem{Alicea} J. Alicea, New directions in the pursuit of Majorana
fermions in solid state systems, Rep. Prog. Phys. \textbf{75,} 076501 (2012).

\bibitem{Pfeuty} P. Pfeuty, The one-dimensional Ising model with a
transverse field, Ann. Phys. (NY) \textbf{57,} 79 (1970).

\bibitem{SachdevBook} S. Sachdev, \textit{Quantum Phase Transitions}
(Cambridge University Press, Cambridge, England, 1999). 

\bibitem{Law} X. J. Liu, C. L. M. Wong, and K. T. Law, Non-abelian majorana
doublets in time-reversal-invariant topological superconductors, Phys. Rev.
X \textbf{4,} 021018 (2014).

\bibitem{Tong} X. H. Wang, T. T. Liu, Y. Xiong, and P. Q. Tong, Spontaneous $%
\mathcal{PT}$-symmetry breaking in non-Hermitian Kitaev and extended Kitaev
models, Phys. Rev. A \textbf{92,} 012116 (2015).

\bibitem{Yuce} C. Yuce, Majorana edge modes with gain and loss, Phys. Rev. A 
\textbf{93,} 062130 (2016).

\bibitem{You} Q. B. Zeng, B. G. Zhu, S. Chen, L. You, and R. L\"{u},
Non-Hermitian Kitaev chain with complex on-site potentials, Phys. Rev. A 
\textbf{94,} 022119 (2016).

\bibitem{Klett} M. Klett, H. Cartarius, D. Dast, J. Main, and G. Wunner,
Relation between $\mathcal{PT}$-symmetry breaking and topologically
nontrivial phases in the Su-Schrieffer-Heeger and Kitaev models, Phys. Rev.
A \textbf{95,} 053626 (2017).

\bibitem{Menke} H. Menke and M. M. Hirschmann, Topological quantum wires
with balanced gain and loss, Phys. Rev. B \textbf{95,} 174506 (2017).

\bibitem{LCPRB} C. Li, X. Z. Zhang, G. Zhang, and Z. Song, Topological
phases in a Kitaev chain with imbalanced pairing, Phys. Rev. B \textbf{97},
115436 (2018).

\bibitem{YXMPRA} X. M. Yang and Z. Song, Resonant generation of a -wave
Cooper pair in a non-Hermitian Kitaev chain at the exceptional point, Phys.
Rev. B \textbf{102,} 022219 (2020).

\bibitem{DMTN} {D. Mondal and T. Nag, Anomaly in the dynamical
quantum phase transition in a non-Hermitian system with extended gapless
phases, Phys. Rev. B \textbf{106}, 054308 (2022).}

\bibitem{ZXZ} X. Z. Zhang and Z. Song, Non-Hermitian anisotropic XY model
with intrinsic rotation-time-reversal symmetry, Phys. Rev. A \textbf{87,}
012114 (2013).

\bibitem{LCPRA90} C. Li, G. Zhang, X. Z. Zhang, and Z. Song, Conventional
quantum phase transition driven by a complex parameter in a non-Hermitian $%
\mathcal{PT}$-symmetric Ising model, Phys. Rev. A \textbf{90,} 012103 (2014).

\bibitem{LCPRA92} C. Li and Z. Song, Finite-temperature quantum criticality
in a complex-parameter plane, Phys. Rev. A \textbf{92,} 062103 (2015).

\bibitem{LCPRA94} C. Li, G. Zhang, and Z. Song, Chern number in Ising models
with spatially modulated real and complex fields, Phys. Rev. A \textbf{94,}
052113 (2016).

\bibitem{Franz} M. Franz, Majorana's wires, Nat. Nanotech. \textbf{8,}
149--152 (2016).

\bibitem{Ueda} Y. Ashida, S. Furukawa, and M. Ueda, Parity-time-symmetric
quantum critical phenomena, Nat. Commun. \textbf{8,} 15791 (2017).

\bibitem{Raghu} F. D. M. Haldane and S. Raghu, Possible realization of
directional optical waveguides in photonic crystals with broken
time-reversal symmetry, Phys. Rev. Lett. \textbf{100,} 013904 (2008).

\bibitem{Shindou} R. Shindou, R. Matsumoto, S. Murakami, and J. Ohe,
Topological chiral magnonic edge mode in a magnonic crystal, Phys. Rev. B 
\textbf{87,} 174427 (2013).

\bibitem{HF} { R. E. Stanton, Hellmann-Feynman Theorem and
Correlation Energies, J. Chem. Phys. \textbf{36}, 1298 (1962).}

\bibitem{ZGPRL} G. Zhang, Z. Song, Topological characterization of extended
quantum Ising models, Phys. Rev. Lett. \textbf{17,} 177204 (2015).

\bibitem{Thouless} D. J. Thouless, Topological Quantum Numbers in
Nonrelativistic Physics (World Scientific, Singapore, 1998).

\bibitem{Kane} M. Z. Hasan and C. L. Kane, Colloquium: Topological
insulators, Rev. Mod. Phys. \textbf{82,} 3045 (2010).

\bibitem{Zhang} X.-L. Qi and S.-C. Zhang, Topological insulators and
superconductors, Rev. Mod. Phys. \textbf{83,} 1057 (2011).

\bibitem{Lu} L. Lu, J. D. Joannopoulos, and M. Solja\v{c}i\'{c}, Topological
photonics, Nat. Photonics \textbf{8,} 821 (2014).

\bibitem{Esaki} K. Esaki, M. Sato, K. Hasebe, and M. Kohmoto, Edge states
and topological phases in non-Hermitian systems, Phys. Rev. B \textbf{84,}
205128 (2011).

\bibitem{Hu} Y. C. Hu and T. L. Hughes, Absence of topological insulator
phases in non-Hermitian $\mathcal{PT}$-symmetric Hamiltonians, Phys. Rev. B 
\textbf{84,} 153101 (2011).

\bibitem{Bender2} C. M. Bender, Making sense of non-Hermitian Hamiltonians,
Rep. Prog. Phys. \textbf{70,} 947--1018 (2007).

\bibitem{Jones} H. F. Jones, On pseudo-Hermitian Hamiltonians and their
Hermitian counterparts, J. Phys. A: Math. Gen. \textbf{38,} 1741-1746 (2005).

\bibitem{Ali6} A. Mostafazadeh, On pseudo-Hermitian Hamiltonians and their
Hermitian counterparts, J. Phys. A: Math. Gen. \textbf{38,} 6557 (2005).

\bibitem{Ali7} A. Mostafazadeh, Metric operator in pseudo-Hermitian quantum
mechanics and the imaginary cubic potential, J. Phys. A: Math. Gen. \textbf{%
39,} 10171 (2006).

\bibitem{Ali8} A. Mostafazadeh, Delta-function potential with a complex
coupling, J. Phys. A: Math. Gen. \textbf{39,} 13495 (2006).

\bibitem{ZKLPRL} K. L. Zhang and Z. Song, Quantum phase transition in a
quantum Ising chain at nonzero temperatures, Phys. Rev. Lett. \textbf{126},
116401 (2021).

\bibitem{ZKLPRB} K. L. Zhang and Z. Song, Ising chain with topological
degeneracy induced by dissipation, Phys. Rev. B \textbf{101}, 245152 (2020).

\bibitem{SW} {S. Weimann, M. Kremer, Y. Plotnik, Y. Lumer, S.
Nolte, K. G. Makris, M. Segev, M. C. Rechtsman and A. Szameit, Topologically
protected bound states in photonic parity-time-symmetric crystals, Nat.
Mater. \textbf{16}, 433438 (2017).}

\bibitem{SYBPRB} {Y. B. Shi, K. L. Zhang, and Z. Song, Dynamic
generation of nonequilibrium superconducting states in a Kitaev chain, Phys.
Rev. B \textbf{106}, 184505 (2022).}
\end{thebibliography}
\end{document}